\documentclass[english,superscriptaddress,reprint,two column]{revtex4-1}
\usepackage[T1]{fontenc}
\usepackage{color}
\usepackage{amsmath}
\usepackage{amssymb}
\usepackage{graphicx}
\usepackage{wasysym}
\usepackage{amssymb}
\usepackage[normalem]{ulem}

\makeatletter
\usepackage{amsfonts}
\usepackage{xcolor}
\usepackage{transparent}

\makeatother

\usepackage{babel}
\begin{document}

\title{Engineering asymmetric steady-state Einstein-Podolsky-Rosen steering in macroscopic hybrid systems}

\author{Xinyao Huang}
\affiliation{State Key Laboratory for Mesoscopic Physics, School of Physics, Peking University, Beijing 100871, China}
\affiliation{Nano-optoelectronics Frontier Center of the Ministry of Education $\&$ Collaborative Innovation Center of Quantum Matter, Peking University, Beijing 100871, China}
\author{Emil Zeuthen}
\affiliation{Niels Bohr Institute, University of Copenhagen, DK-2100 Copenhagen, Denmark}
\author{Qihuang Gong}
\affiliation{State Key Laboratory for Mesoscopic Physics, School of Physics, Peking University, Beijing 100871, China}
\affiliation{Nano-optoelectronics Frontier Center of the Ministry of Education $\&$ Collaborative Innovation Center of Quantum Matter, Peking University, Beijing 100871, China}
\affiliation{Beijing Academy of Quantum Information Sciences, Haidian District, Beijing 100193, China}
\affiliation{Collaborative Innovation Center of Extreme Optics, Shanxi University, Taiyuan, Shanxi 030006, China}
\author{Qiongyi He}
\affiliation{State Key Laboratory for Mesoscopic Physics, School of Physics, Peking University, Beijing 100871, China}
\affiliation{Nano-optoelectronics Frontier Center of the Ministry of Education $\&$ Collaborative Innovation Center of Quantum Matter, Peking University, Beijing 100871, China}
\affiliation{Beijing Academy of Quantum Information Sciences, Haidian District, Beijing 100193, China}
\affiliation{Collaborative Innovation Center of Extreme Optics, Shanxi University, Taiyuan, Shanxi 030006, China}

\begin{abstract}
Generation of quantum correlations between separate objects is of significance both in fundamental physics and in quantum networks. One important challenge is to create the directional ``spooky action-at-a-distance'' effects that Schr\"{o}dinger called ``steering'' between two macroscopic and massive objects.
Here, we analyze a generic scheme for generating steering correlations in cascaded hybrid systems in which two distant oscillators with effective masses of opposite signs are coupled to a unidirectional light field,
a setup which is known to build up quantum correlations by means of quantum back-action evasion. % in Einstein-Podolsky-Rosen (EPR) variables of the system.
The unidirectional coupling of the first to the second oscillator via the light field 
can be engineered to enhance steering in both directions 
and provides an active method for controlling the asymmetry of steering. %within the apparatus itself instead of by adding asymmetric amounts of noise or loss to each subsystem. 
We show that the resulting scheme can efficiently generate unconditional steady-state Einstein-Podolsky-Rosen steering between the two subsystems, even in the presence of thermal noise and optical losses. 
As a scenario of particular technological interest in quantum networks, we use our scheme to engineer enhanced steering from an untrusted node with limited tunability (in terms of interaction strength and type with the light field) to a trusted, highly tunable node, hence offering a path to implementing one-sided device-independent quantum tasks.

\end{abstract}
\maketitle

\section{INTRODUCTION} 

The observation of the ``spooky action-at-a-distance'' effects predicted in the famous Einstein-Podolsky-Rosen (EPR) paradox~\cite{Einstein1935} is one of the most basic tests of quantum correlations. This effect was termed ``steering'' by Schr\"{o}dinger~\cite{Schrodinger1935}.
It describes that two spatially separated systems (which are held by Alice and Bob) share an entangled state and the local measurement on one system can instantaneously affect (steer) the state of the other system~\cite{Schrodinger1936}. In 2007, EPR steering was rigorously defined in terms of violations of a local hidden state model by Wiseman {\it et al.}~\cite{Wiseman2007,Jones2007}. As a class of quantum correlations, steering is a strict subset of entanglement in terms of inseparability and a strict superset of Bell nonlocality~\cite{Wiseman2007,Quintino2015}. Moreover, steering is a directional form of nonlocality; in particular, in some scenarios one may have one-way steering, where only Alice can steer Bob, but not vice versa~\cite{Wiseman2007,Quintino2015,Handchen2012}. This is distinct from entanglement and Bell nonlocality, which are shared symmetrically between the systems.  

In addition to being of fundamental interest, steering is also important to quantum networks as it provides a way to verify entanglement without requiring trust of the equipment at each node of the network; e.g., if Alice can be convinced that Bob can steer the state of her system  by local measurements on his own system, then entanglement between them has been confirmed without trusting Bob's device~\cite{Wiseman2007,Jones2007,Cavalcanti2009, Reid2009, Cavalcanti2017}. This is referred to as a one-sided (1s) device-independent (DI) protocol, an intermediate framework between the standard device-dependent approach based on state inseparability~\cite{Gisin2002} and the fully DI protocol~\cite{Acin2007} based on loophole-free Bell tests. Considering the extraordinarily challenging character of the fully DI experiments, 1sDI protocols based on steering offer a more feasible approach to performing quantum secure tasks on a network where reliability or tampering of devices, dishonest observers, etc.~could be an issue, e.g., 1sDI quantum key distribution~\cite{Branciard2012,Walk2016,Gehring2015}, quantum secret sharing~\cite{Armstrong2015,Xiang2017,Kogias2017}, and quantum teleportation~\cite{He2015,Reid2013,Chiu2016}.

Motivated both by the fundamental interest and applications, observing EPR steering in various systems and uncovering the potential of 1sDI protocols using steering as a resource~\cite{Gallego2015} have attracted a great deal of attention.
Experimental demonstrations of EPR steering have to date mostly been realized with optical fields~\cite{Reid2009,Saunders2010, Handchen2012, Steinlechner2013,Schneeloch2013, Moreau2014,Sun2014, Cavalcanti2015,Li2015,Armstrong2015,Sun2016,  Wollmann2016, Xiao2017, Deng2017,Qin2017,Tischler2018}. Some of these experiments have shown asymmetric two-way steering~\cite{Sun2016} and even one-way steering~\cite{Handchen2012,Armstrong2015, Wollmann2016,Xiao2017,Deng2017, Qin2017,Tischler2018} by producing asymmetric states via the addition of variable loss or external thermal noise to one of the subsystems. Recently, however, EPR steering has also been observed in massive, macroscopic systems such as a Bose-Einstein condensate in spin degrees of freedom~\cite{Peise2015}, between an atomic ensemble and a photon~\cite{Zhang2019,Dabrowski2018}, and between separated Bose-Einstein condensates~\cite{Fadel2018,Kunkel2018}.
%In light of these exceptional developments, a further question is raised: how to demonstrate EPR-steering in macroscopic hybrid systems ?
Although the above developments represent progress towards the goal of establishing steering between nodes in a quantum network, the question remains how to realize this over appreciable distances and between potentially disparate systems as envisioned in future hybrid quantum networks~\cite{Kurizki2015}.

In this paper, we present a scheme that deals with these challenges.
It assumes that the two subsystems, between which correlations are to be generated, are connected by a propagating unidirectional light field traveling in an optical fiber (or other turbulence-free channel), a natural scenario in a quantum network.
It is well known that if the two subsystems behave as harmonic oscillators with effective masses of opposite signs, then quantum correlations between them can be generated due to quantum back-action evasion in the EPR variables of the joint system~\cite{Hammerer2009,Hammerer2010,Tsang2010,Muschik2011,Tsang2012,Vasilyev2013,Woolley2014}. To date, this mechanism has been successfully applied to generate entanglement between two similar objects~\cite{Julsgaard2001, Krauter2011,Ockeloen-Korppi2018} and to cancel quantum back action in a hybrid spin-optomechanical system~\cite{Moller2017}. Very recently, it was shown how this mechanism can be engineered to generate unconditional steady-state EPR entanglement in cascaded hybrid systems, with a performance  matching that of comparable conditional schemes~\cite{Huang2018}. 
The present paper extends the entanglement scheme in Ref.~\cite{Huang2018} to the EPR-steering scenario and shows
how this allows controlling the steering asymmetry between the two directions by tuning the light-matter interaction strengths and types. 
This provides an effective method for controlling the asymmetry of steering within the apparatus as opposed to the typical approach of adding asymmetric amounts of external noise or loss.
To show its potential application in quantum networks, we demonstrate the ability of our scheme to engineer enhanced  steering from an untrusted node to a trusted node, %where "cheap/expensive" refers to the tunability of the node in regards to the interaction strength and type with the optical field.
thereby making 1sDI quantum information tasks between separate macroscopic systems feasible under realistic conditions. %in a network where, e.g., numerous "cheap" nodes are connected to a central "expensive" node.

The paper is organized as follows: In Sec.~\ref{sec: model}, we describe the mathematical model for the cascaded hybrid system and then, in Sec.~\ref{sec: EPRproduct}, introduce the criteria for EPR steering. In Sec.~\ref{sec: Asymmetry}, we demonstrate how to generate and control the asymmetry of steady-state EPR steering by engineering asymmetric couplings to the light field. Next, in Sec.~\ref{sec: Optimization}, we consider optimal steering within our scheme, %present the finding that the optimal asymmetric couplings can enhance the steerabilities, 
with particular focus on improving the generation of steering from a subsystem with limited tunability to a freely tunable subsystem. Finally, we discuss the optimality and implementation of our scheme in Sec.~\ref{sec: discussion} and summarize our results in Sec.~\ref{sec: conclusion}.

\section{Model} \label{sec: model}

\begin{figure}[!h]
\centering{
     \includegraphics[width=0.98\columnwidth]{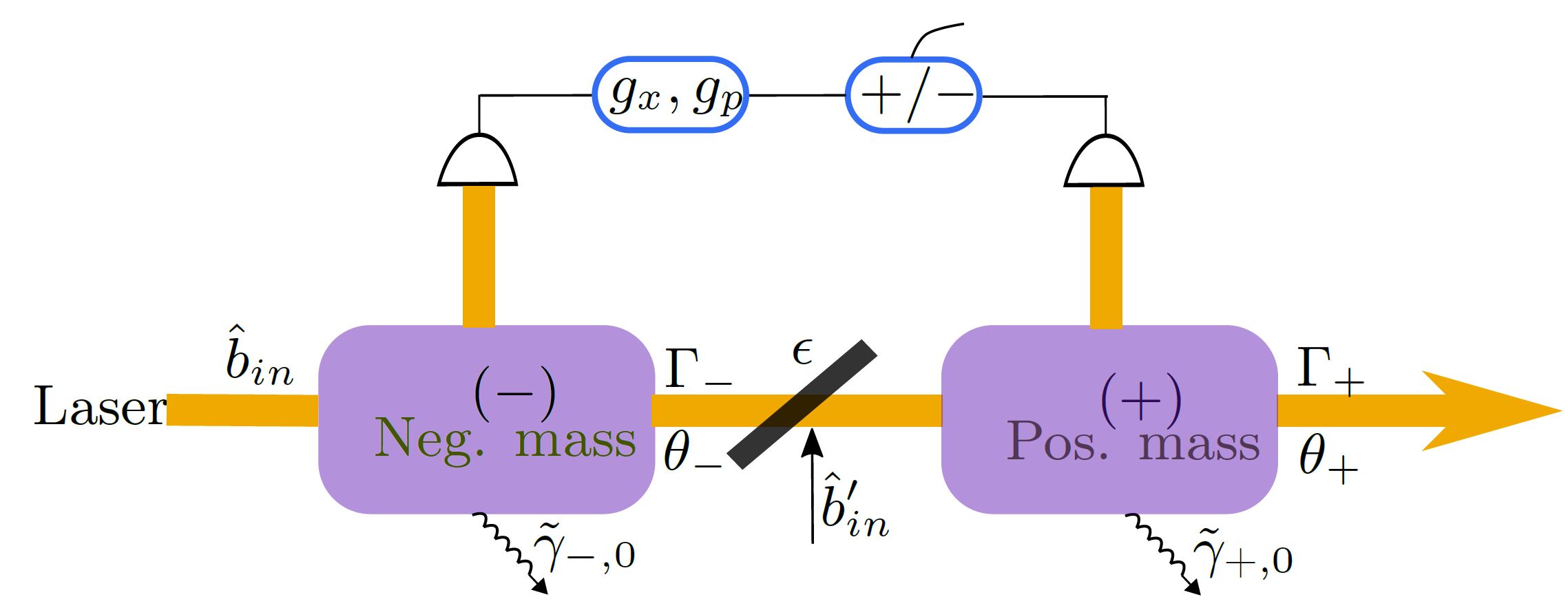}}
\caption{A cascaded hybrid system composed of two oscillators with effective masses of opposite signs and individual decoherence rates $\tilde{\gamma}_{j,0}$, $j\in\{+,-\}$, coupled with a unidirectional propagating light field. The light-oscillator interaction is induced by a coherent laser drive (assumed to be quantum limited) and is parametrized by the interaction strength $\Gamma_{j}$ and angle $\theta_{j}$ controlling the interaction type. Additionally, the second oscillator is driven by an uncorrelated vacuum field $\hat{b}_{in}'$ due to the presence of the transmission (power) loss $\epsilon$ between the subsystems.
In order to verify EPR steering, independent local measurements are made on the individual subsystems (top part of figure). These are combined using gain factors $g_{x,p}$, which can be optimized to minimize the product of the uncertainties (standard deviations) $\Delta(\hat{X}_{+}-g_x \hat{X}_{-})$ and $\Delta(\hat{P}_{+}+g_p \hat{P}_{-})$ entering the steering criteria.}
\label{fig:set up}
\end{figure}

We first briefly review the model introduced in Ref.~\cite{Huang2018}, where two spatially separated oscillators with effective masses of opposite signs, $\text{sgn}(m_{+})=-\text{sgn}(m_{-})$, are interacting with a propagating light field in cascade as shown in Fig.~\ref{fig:set up}.  The coupling with their individual thermal reservoirs induces the respective thermal decoherence rates $\tilde{\gamma}_{j,0}$, $j\in\{+,-\}$, of the two oscillators. When we employ the bosonic annihilation and creation operators $\hat{a}_{j}$ and $\hat{a}^{\dagger}_{j}$ with commutator $[\hat{a}_{j},\hat{a}^{\dagger}_{j}]=1$ to describe the localized oscillators,  the corresponding dimensionless canonical operators are $\hat{X}_{j}=(\hat{a}_{j}+\hat{a}_{j}^{\dagger})/\sqrt{2}$ and $\hat{P}_{j}=(\hat{a}_{j}-\hat{a}_{j}^{\dagger})/(\sqrt{2}i)$, normalized to their zero-point  position $x_{j,\text{zpf}}=\sqrt{\hbar/(|m_{j}|\omega_{j})}$ and momentum $p_{j,\text{zpf}}=\hbar/x_{j,\text{zpf}}$ amplitudes, where $\omega_{j}$ is the resonance frequency of the oscillator. In this representation, the sign of the oscillator mass $m_j$ is absorbed into the effective resonance frequency $\Omega_{j}=\text{sgn}(m_{j})\omega_{j}$, so that a negative mass amounts to a negative effective resonance frequency. Such a negative effective resonance frequency can emerge in systems prepared in the vicinity of an energetically maximal state, 
as can be implemented, e.g., in a macroscopic collective spin degree of freedom by preparing an inverted spin population~\cite{Julsgaard2001,Hammerer2010, Muessel2014, McConnell2015}. The same effective Hamiltonian of a negative-mass oscillator can also be realized in optomechanical systems~\cite{Aspelmeyer2014} by applying two-tone driving schemes~\cite{Tan2013, Woolley2013, Woolley2014} (Refs.~\cite{Woolley2013,Woolley2014} have been implemented electromechanically~\cite{Ockeloen-Korppi2016,Ockeloen-Korppi2018}).

Considering a general quadratic interaction between the unidirectional optical field and the oscillators, the dynamics of the hybrid system can be described by the Hamiltonian \cite{Muschik2011,Vasilyev2013} ($\hbar=1$)
\begin{equation}
\hat{H}=\hat{H}_{\text{sys}}+\hat{H}_{\text{field}}+\hat{H}_{\text{int}}+\hat{H}_{\text{diss}},
\label{eq:H}
\end{equation}
where
\begin{align}
\hat{H}_{\text{sys}}={}& \sum_{j\in\{+,-\}} \frac{\Omega_j}{2}(\hat{X}^2_j + \hat{P}^2_j),
\nonumber\\
\hat{H}_{\text{field}}={}&\int_{-\infty}^{\infty}\Omega\hat{b}^{\dagger}(\Omega)\hat{b}(\Omega)d\Omega,
\nonumber\\
\hat{H}_{\text{int}} ={}&\sqrt{\Gamma_{-B}}\hat{a}_{-}^{\dagger}\hat{b}(t_-)+\sqrt{\Gamma_{-P}}\hat{a}_{-}^{\dagger}\hat{b}^{\dagger}(t_-)
\nonumber\\
&\sqrt{\Gamma_{+B}}\hat{a}_{+}^{\dagger}\hat{b}(t_+)+\sqrt{\Gamma_{+P}}\hat{a}_{+}^{\dagger}\hat{b}^{\dagger}(t_+)+\text{H.c.}
\label{eq:int_H}
\end{align}
Here the optical vacuum field is described by the operator $\hat{b}(t)\equiv \hat{b}_{l}(t)+\hat{b}_{u}(t) = (2\pi)^{-1/2}(\int_{-\infty}^{0}+\int_{0}^{\infty})\hat{b}(\Omega)e^{-i\Omega t}d\Omega$ (in the frame rotating at the drive laser frequency $\omega_{\text{L}}$), which is decomposed into upper and lower sideband operators $b_{u}(t)$, $b_{l}(t)$ in the time domain. The oscillator-field interaction comprises both a beam-splitter-type interaction $(\hat{a}_{j}^{\dagger}\hat{b}+\text{H.c.})$, which can generate a state transfer between the oscillator and the optical mode, and a parametric-down-conversion-type interaction $(\hat{a}_{j}^{\dagger}\hat{b}^{\dagger}+\text{H.c.})$, known to create squeezing and entanglement between two modes. The coefficients of the two interaction types can be parametrized as $\Gamma_{jB}=\Gamma_{j}\sin^{2}\theta_{j}$, $\Gamma_{jP}=\Gamma_{j}-\Gamma_{jB}$ in terms of the interaction strength $\Gamma_{j}$ and the interaction angle $\theta_{j} \in \{0,\pi/2\}$ representing the relative magnitude of the two processes, thereby controlling the type of the oscillator-field interaction. For instance when $\theta_{j}=\pi/4$, the so-called quantum nondemolition (QND) interaction is achieved (the two processes have equal weights). Finally, the dissipation Hamiltonian $\hat{H}_{\text{diss}}$ accounts for the decoherence processes due to the coupling of the oscillators with their individual thermal baths %, leading to decoherence rates $ \tilde{\gamma}_{\pm,0}$, 
and the extraneous vacuum field $\hat{b}'_{\text{in}}$ in the presence of  transmission loss $\epsilon > 0$.

We assume $|\Omega_{+}|=|\Omega_{-}|$ to match the two oscillators and, for specificity, $t_-<t_+$ for the cascade ordering, as illustrated in Fig.~\ref{fig:set up} (note that it is unimportant which of the two subsystems implements the negative mass oscillator). The unidirectional light field arriving at the second ($+$) subsystem then satisfies $\hat{b}_{+,\text{in}}=e^{i\phi}\sqrt{1-\epsilon}\hat{b}_{-,\text{out}}+\sqrt{\epsilon}\hat{b}'_{\text{in}}$. Here $\hat{b}_{j,\text{in/out}}$ is the input/output quantum field of subsystem $j\in\{+,-\}$, and $\hat{b}'_{\text{in}}$ represents the uncorrelated vacuum noise due to losses that drives only the second subsystem, as depicted in Fig.~\ref{fig:set up}. $\phi$ is an adjustable quadrature rotation phase factor  between the two subsystems, which we optimize numerically to minimize the EPR-steering parameter.

By performing the rotating-wave approximation (RWA) in the regime $|\Omega_{\pm}| \gg \Gamma_{\pm} \gtrsim \tilde{\gamma}_{\pm,0}$ and choosing the optimal quadrature $\phi=0$, the Langevin equations after 
elimination of the optical field read
\begin{align}
\frac{d}{d t}\hat{a}_{-}(t)  ={} & i(\sqrt{\Gamma_{-B}}\hat{b}_{l,\text{in}}+\sqrt{\Gamma_{-P}}\hat{b}_{u,\text{in}}^{\dagger})
\nonumber\\
& -\frac{\gamma_{-}}{2}\hat{a}_{-}+\sqrt{\gamma_{-,0}}\hat{a}_{-,\text{in}},
\nonumber \\
\frac{d}{d t}\hat{a}_{+}(t)  = {}& i\sqrt{1-\epsilon}(\sqrt{\Gamma_{+B}}\hat{b}_{u,\text{in}}+\sqrt{\Gamma_{+P}}\hat{b}_{l,\text{in}}^{\dagger})
\nonumber\\
&+i\sqrt{\epsilon}(\sqrt{\Gamma_{+B}}\hat{b}'_{u,\text{in}}+\sqrt{\Gamma_{+P}}\hat{b}'^{\dagger}_{l,\text{in}})
\nonumber\\
&-\frac{\gamma_{+}}{2}\hat{a}_{+}+\sqrt{\gamma_{+,0}}\hat{a}_{+,\text{in}}+\sqrt{1-\epsilon}R\hat{a}_{-}^{\dagger},
\label{eq:Le-nophase}
\end{align}
where the total damping rate $\gamma_{\pm}=\gamma_{\pm,0}+\gamma_{\pm,\text{opt}}$ includes the intrinsic damping rate $\gamma_{\pm,0}$ and the dynamical optical broadening $\gamma_{\pm,\text{opt}} \equiv \Gamma_{\pm B}-\Gamma_{\pm P}$. Dynamical stability requires $\gamma_{\pm}>0$ and we will assume this condition to be fulfilled throughout the present analysis.
Both oscillators are driven by the common optical vacuum input field $\hat{b}_{u/l, \text{in}}$ which has zero thermal occupation, i.e., $\langle\hat{b}_{u/l,\text{in}}(t) \hat{b}_{u/l,\text{in}}^{\dagger}(t')\rangle=\delta(t-t')$, as does the extraneous vacuum field $\langle\hat{b}'_{u/l,\text{in}}(t) \hat{b}'^{\dagger}_{u/l,\text{in}}(t')\rangle=\delta(t-t')$. In addition, both oscillators are coupled to their individual thermal reservoirs with expectation values given by $\langle\hat{a}_{\pm, \text{in}}(t)\hat{a}_{\pm, \text{in}}^{\dagger}(t')\rangle=(\bar{n}_{\pm}+1)\delta(t-t')$, where $\bar{n}_{\pm}$ are the mean thermal occupation numbers, inducing the thermal decoherence rates $\tilde{\gamma}_{\pm,0}\equiv \gamma_{\pm,0}(\bar{n}_{\pm,0}+1/2)$.
In terms of these, we introduce the quantum cooperativities $C_{j}\equiv\Gamma_{j}/\tilde{\gamma}_{j,0}$ of the subsystems, i.e., the ratio of the coherent coupling and thermal decoherence rates, which will play a prominent role in the analysis below. Note that this definition of quantum cooperativity in terms of the rate $\Gamma_j = \Gamma_{jB}+\Gamma_{jP}$ differs from that typically used in optomechanics~\cite{Aspelmeyer2014} (except in the resolved-sideband regime).

The unidirectional light field will (in general) read out the response of the first oscillator $(-)$ and map this information to the second oscillator $(+)$, thereby inducing an effective directional coupling of the first to the second oscillator with the rate $R=\sqrt{\Gamma_{-B}\Gamma_{+P}}-\sqrt{\Gamma_{+B}\Gamma_{-P}}=-\sqrt{\Gamma_+\Gamma_-}\sin(\theta_+-\theta_-)$. This directional coupling arises whenever the interaction angles of the two oscillators differ from each other $\theta_+ \neq \theta_-$, so that $R \neq 0$. 
This coupling provides an additional mechanism for building correlations 
(in addition to the common optical bath) 
that allows us to control the asymmetry of EPR steering and enhance the steerabilities, as will be explored below.

Integrating Eqs.~\eqref{eq:Le-nophase} from the initial time $t_0=-\infty$ to $t$ leads to solutions from which the variances and covariances %(i.e., the quantum fluctuations around the mean value) 
of the quadratures operators $\hat{X}_\pm$, $\hat{P}_\pm$ (defined relative to the classical amplitudes, respectively) can be evaluated. These steady-state values within the RWA are~\cite{Huang2018}
\begin{align}
&(\Delta\hat{X}_-)^2=(\Delta\hat{P}_-)^2=\frac{\Gamma_{-}/2+\tilde{\gamma}_{-,0}}{\gamma_{-}},
\nonumber\\
&(\Delta\hat{X}_{+})^2=(\Delta\hat{P}_{+})^2=\frac{\Gamma_{+}/2+\tilde{\gamma}_{+,0}+2\sqrt{1-\epsilon}R\langle \hat{X}_{+},\hat{X}_{-}\rangle}{\gamma_{+}},
\nonumber\\
&\langle \hat{X}_{+},\hat{X}_{-}\rangle =-\langle \hat{P}_{+},\hat{P}_{-}\rangle\nonumber\\
&=\frac{-\sqrt{1-\epsilon}\left[\sqrt{\Gamma_{+}\Gamma_{-}}\sin(\theta_{+}+\theta_{-})-2R(\Delta\hat{X}_-)^2\right]}{\gamma_{+}+\gamma_{-}},
\label{eq:ss_EPRvar}
\end{align}
where we use the notations $(\Delta \hat{x})^{2}\equiv \langle \hat{x}^{2}\rangle- \langle \hat{x} \rangle ^{2}$ and $\langle\hat{x},\hat{y}\rangle \equiv(\langle \hat{x}\hat{y}\rangle +\langle \hat{y}\hat{x}\rangle)/2 - \langle \hat{x} \rangle  \langle\hat{y} \rangle$. 
%\textcolor{blue}{Note that the operators $\hat{X}_\pm$, $\hat{P}_\pm$ have zero mean and represent the quantum fluctuations around a classical fixed point.}

\section{EPR-steering Criterion} \label{sec: EPRproduct}

We now present the criterion for having EPR-steering correlations in our system.
The presence of EPR steering can be detected by adopting Reid's EPR criterion~\cite{Reid1989,Reid2009}, which is expressed in terms of the product of the inferred uncertainties for the quadrature variables. EPR steering from the first $(-)$ to the second $(+)$ oscillator is confirmed if
\begin{equation}
E_{+\mid -}=\Delta_{\text{inf}}\hat{X}_{+} \Delta_{\text{inf}}\hat{P}_{+}<\frac{1}{2}.
\label{eq:EPRMS}
\end{equation}
Here $\Delta_{\text{inf}}\hat{X}_{+}\equiv\Delta(\hat{X}_{+}-g_x \hat{X}_{-})$ and $\Delta_{\text{inf}}\hat{P}_{+}\equiv\Delta(\hat{P}_{+}+g_p \hat{P}_{-})$ are the inferred uncertainties in the prediction for the values of the second $(+)$ oscillator's position and momentum based on the outcomes of the local measurements performed on the first $(-)$ oscillator, where $g_x$ and $g_p$ are arbitrary real constants, adjusted to minimize the inferred uncertainties. For Gaussian states and measurements, as is the case here, Eq.~\eqref{eq:EPRMS} is necessary and sufficient to detect EPR steering~\cite{Wiseman2007,Reid2009}. The smallness of the steering parameter $E_{+ \mid -}$ gives a measure of the Gaussian steerability from the first $(-)$ to the second $(+)$ subsystem. Ideally it becomes zero when the two oscillators are in a perfect EPR state where the position and momentum of  the second $(+)$ oscillator can be predicted with $100\%$ accuracy, based on the measurements performed on the first $(-)$ oscillator.

From Eqs.~(\ref{eq:ss_EPRvar}), we have the optimal value $g_x=g_p=\langle \hat{X}_{+},\hat{X}_{-}\rangle / (\Delta\hat{X}_{-})^2$ within the RWA, obtained by the variation method $\partial \Delta_{\text{inf}}\hat{X}_{+}(\hat{P}_{+}) /\partial g_x(g_p)=0$. Hence, the minimized inferred quadrature variances are $(\Delta_{\text{inf}}\hat{X}_{+})^2=(\Delta\hat{X}_{+})^2-\langle \hat{X}_{+},\hat{X}_{-}\rangle^{2}/(\Delta\hat{X}_{-})^2$, $(\Delta_{\text{inf}}\hat{P}_{+})^2=(\Delta\hat{P}_{+})^2-\langle \hat{P}_{+},\hat{P}_{-}\rangle^{2}/(\Delta\hat{P}_{-})^2=(\Delta_{\text{inf}}\hat{X}_{+})^2$, such that the steering figure of merit~\eqref{eq:EPRMS} minimized within the RWA reads
\begin{equation}
E_{+\mid -}=(\Delta_{\text{inf}}\hat{X}_{+})^2= (\Delta\hat{X}_{+})^2-\langle \hat{X}_{+},\hat{X}_{-}\rangle^{2}/(\Delta\hat{X}_{-})^2.
\label{eq:EPRMSmin}
\end{equation}
Note that the inferred quadrature variance~\eqref{eq:EPRMSmin} are composed of two terms: The first is the noise in the subsystem to be steered, and the second is the uncertainty reduction achieved by measuring the other, correlated subsystem. 

The criterion for EPR steering in the opposite direction, i.e., the first $(-)$ oscillator is steered by the measurements performed on the second $(+)$ oscillator, follows from swapping the $+$ and $-$ subscripts in the above expressions, yielding
%
%\begin{equation}
%E_{-\mid +}=\Delta_{\text{inf}}\hat{X}_{-} %\Delta_{\text{inf}}\hat{P}_{-}<\frac{1}{2},
%\label{eq:EPRSM}
%\end{equation}
%
%where the inferred uncertainties $\Delta_{\text{inf}}\hat{X}_{-}$ and $\Delta_{\text{inf}}\hat{P}_{-}$ are defined similarly as above, and the steering parameter is minimized to
\begin{equation}
 E_{-\mid +}=(\Delta_{\text{inf}}\hat{X}_{-})^2=(\Delta\hat{X}_{-})^2-\langle \hat{X}_{+},\hat{X}_{-}\rangle^{2}/(\Delta\hat{X}_{+})^2.
 \label{eq:EPRSMmin}
\end{equation}
The system exhibits \emph{two-way steering} when simultaneously $E_{+\mid -}<1/2$ and $E_{-\mid +}<1/2$. If only one of them is satisfied, it is referred to as \emph{one-way steering}. \\

\section{Asymmetric Steering Control} \label{sec: Asymmetry}

We first study the distinctive feature of EPR steering, i.e., the asymmetric correlation distribution between the two subsystems. From Eqs.~\eqref{eq:ss_EPRvar}, we can write down the EPR-steering parameters~\eqref{eq:EPRMSmin} and~\eqref{eq:EPRSMmin} as
\begin{align}
E_{+\mid -}&=(\Delta\hat{X}_{+0})^2-\langle \hat{X}_{+},\hat{X}_{-}\rangle^{2}[\frac{1}{(\Delta\hat{X}_{-0})^2}-\frac{\sqrt{1-\epsilon}f}{\langle \hat{X}_{+},\hat{X}_{-}\rangle}],
\nonumber \\
E_{-\mid +}&=(\Delta\hat{X}_{-0})^{2}-\frac{\langle \hat{X}_{+},\hat{X}_{-}\rangle^{2}}{(\Delta\hat{X}_{+0})^{2}+\sqrt{1-\epsilon}f\langle \hat{X}_{+},\hat{X}_{-}\rangle}, 
\label{eq:E}
\end{align}
where $f=2R/\gamma_{+}$ is a parameter associated with the additional noise interference due to the induced directional coupling. In Eqs.~(\ref{eq:E}), we have used the substitution $(\Delta \hat{X}_+)^2 = (\Delta\hat{X}_{+0})^2 + \sqrt{1-\epsilon}f\langle \hat{X}_{+},\hat{X}_{-}\rangle$, where
\begin{equation}
(\Delta\hat{X}_{+0})^2 \equiv (\Gamma_{+}/2+\tilde{\gamma}_{+,0})/\gamma_{+}     
\label{eq:bare-var}
\end{equation}
represents the bare variance of the second $(+)$ oscillator, which has the same form as $(\Delta\hat{X}_{-})^2 \equiv (\Delta\hat{X}_{-0})^2$ [Eq.~\eqref{eq:ss_EPRvar}], composed only by the light back action and its thermal noise, and is thus independent of the correlations between the subsystems. 
From Eqs.~(\ref{eq:E}), we see that the steering parameters include two parts: the bare variance of each subsystem $(\Delta \hat{X}_{\pm0})^2$ and the noise reduction $\propto \langle \hat{X}_{+},\hat{X}_{-}\rangle^2$ due to the correlations  between the subsystems.
When the two subsystems are coupled to the optical field with the same interaction angles  ($\theta_+=\theta_-$, i.e., $f=0$), the asymmetry of steering in the steady state solely depends on the difference of the bare variances in the two subsystems, which are determined by their individual cooperativities $C_{\pm}$ and decoherence properties ($\gamma_{\pm,0}$, $\bar{n}_{\pm,0}$). Once the interaction types of the two subsystems with the optical field become asymmetric ($\theta_{+}\neq \theta_{-}$, i.e., $f\neq0$), the induced directional coupling of the first $(-)$ to the second $(+)$ subsystem  will affect both the covariance $\langle \hat{X}_{+},\hat{X}_{-}\rangle$ and the quadrature variance $(\Delta \hat{X}_+)^2 = (\Delta\hat{X}_{+0})^2 + \sqrt{1-\epsilon}f\langle \hat{X}_{+},\hat{X}_{-}\rangle$ of the second $(+)$ subsystem, giving rise to the additional nontrivial asymmetry of the two steering parameters $E_{+\mid -}$ and $E_{-\mid +}$ manifest in Eqs.~(\ref{eq:E}); this offers an extra, tunable channel to asymmetrically control EPR steering between the two subsystems within the scheme. 

To study the asymmetric steering that can be engineered via this additional channel,
 we will in the following often assume identical decoherence parameters ($\gamma_{\pm,0}=\gamma_{0}, \bar{n}_\pm=\bar{n}$) of the subsystems to exclude the asymmetry brought about by differing thermal parameters. 
We will show how to efficiently control the degrees and directions of EPR steering by tuning the cooperativities $C_{\pm}$ and interaction types $\theta_\pm$ of each subsystem with regard to the unidirectional optical field.

\subsection{Tuning quantum cooperativities} 

First, we show how to engineer the asymmetry of EPR steering by tuning the quantum cooperativities $C_\pm$ of each subsystem for different choices of fixed interaction angles, taking $\theta_-=0.35\pi$ and considering the three cases $\theta_+=0.35\pi$ ($f=0$), $\theta_+=0.3\pi$ ($f>0$), and $\theta_+=0.4\pi$ ($f<0$), respectively.

\begin{figure}[!h]
\centering{
     \includegraphics[width=0.48\columnwidth]{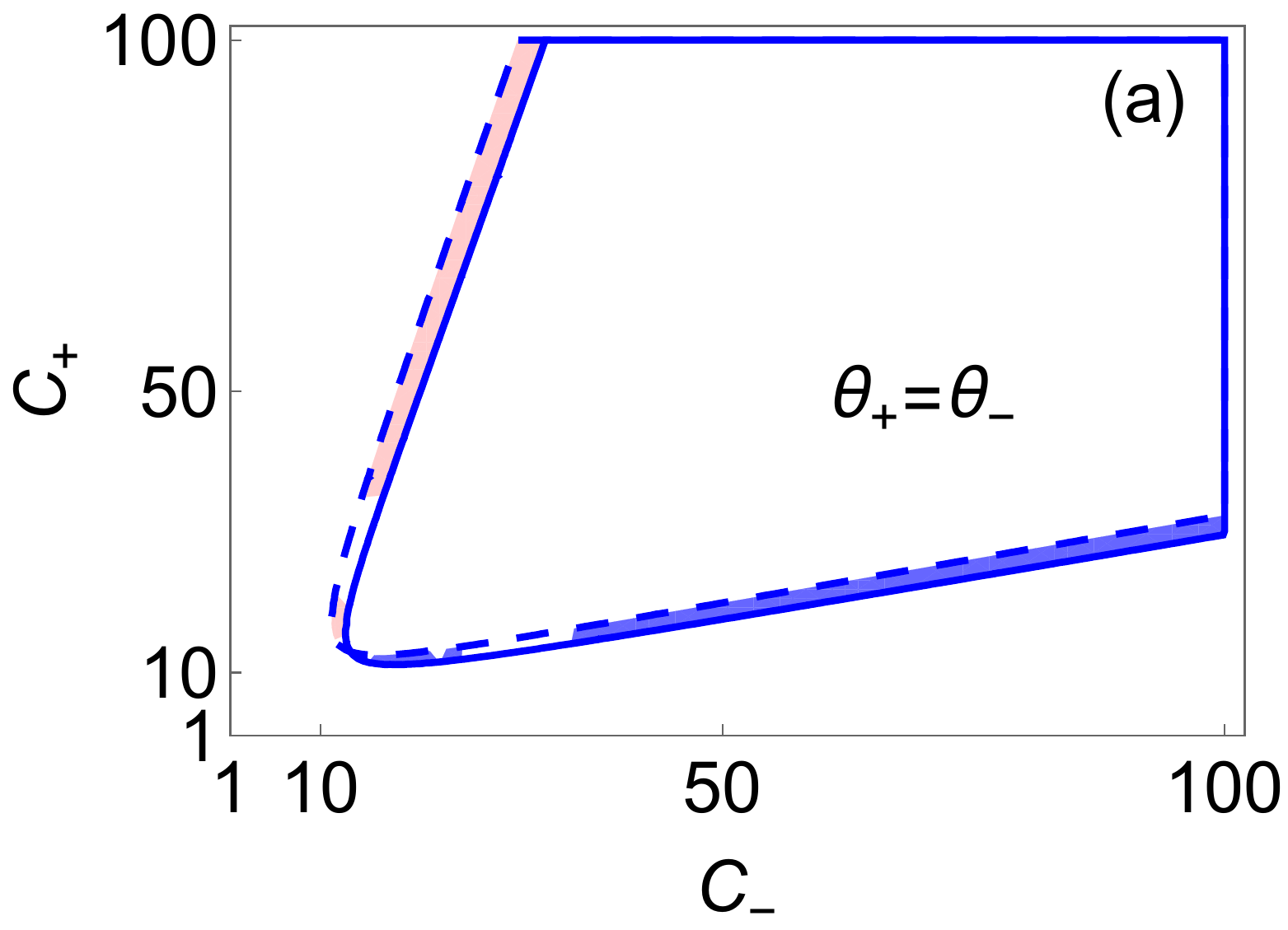}
  \includegraphics[width=0.46\columnwidth]{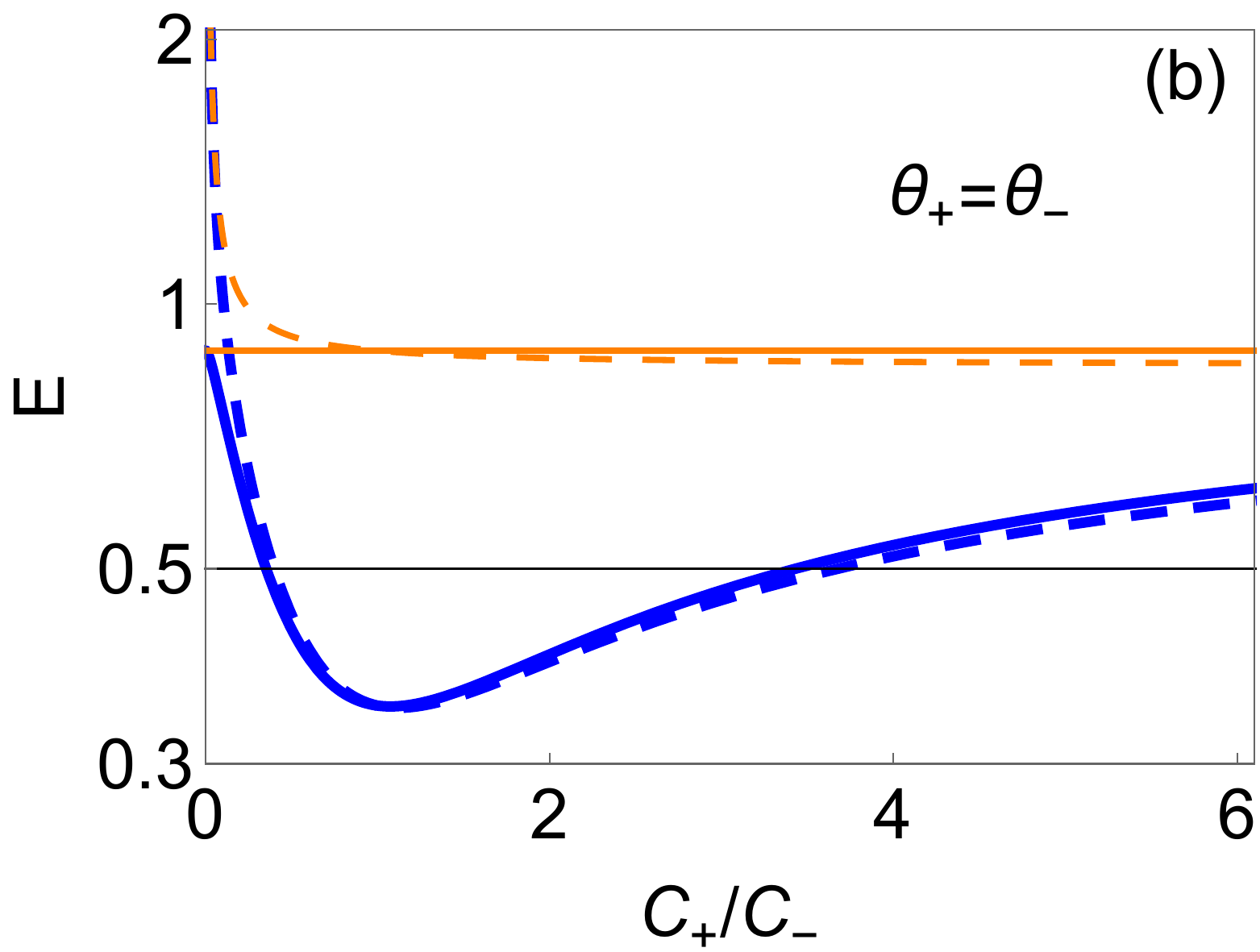}
\includegraphics[width=0.48\columnwidth]{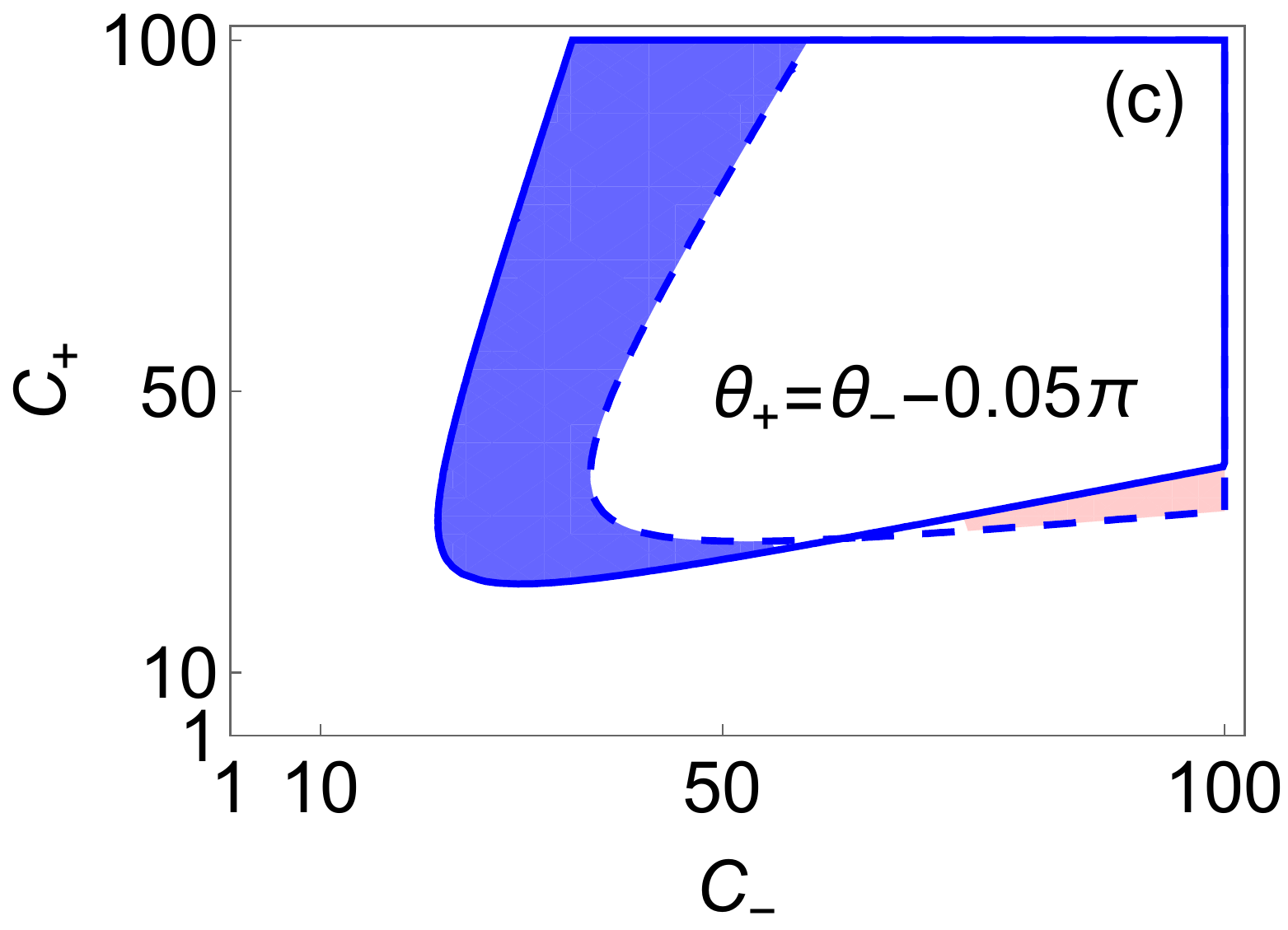}
\includegraphics[width=0.46\columnwidth]{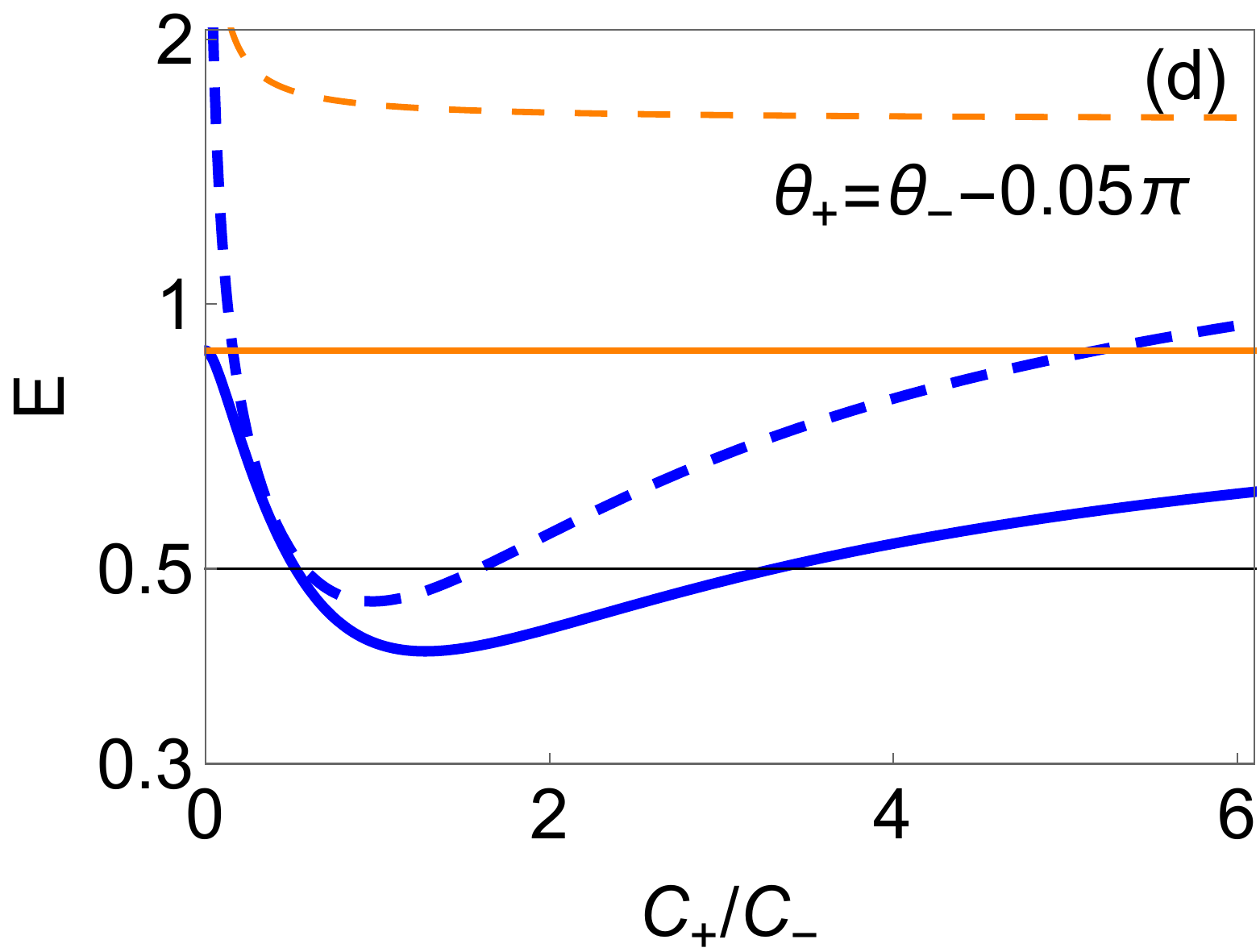}
\includegraphics[width=0.48\columnwidth]{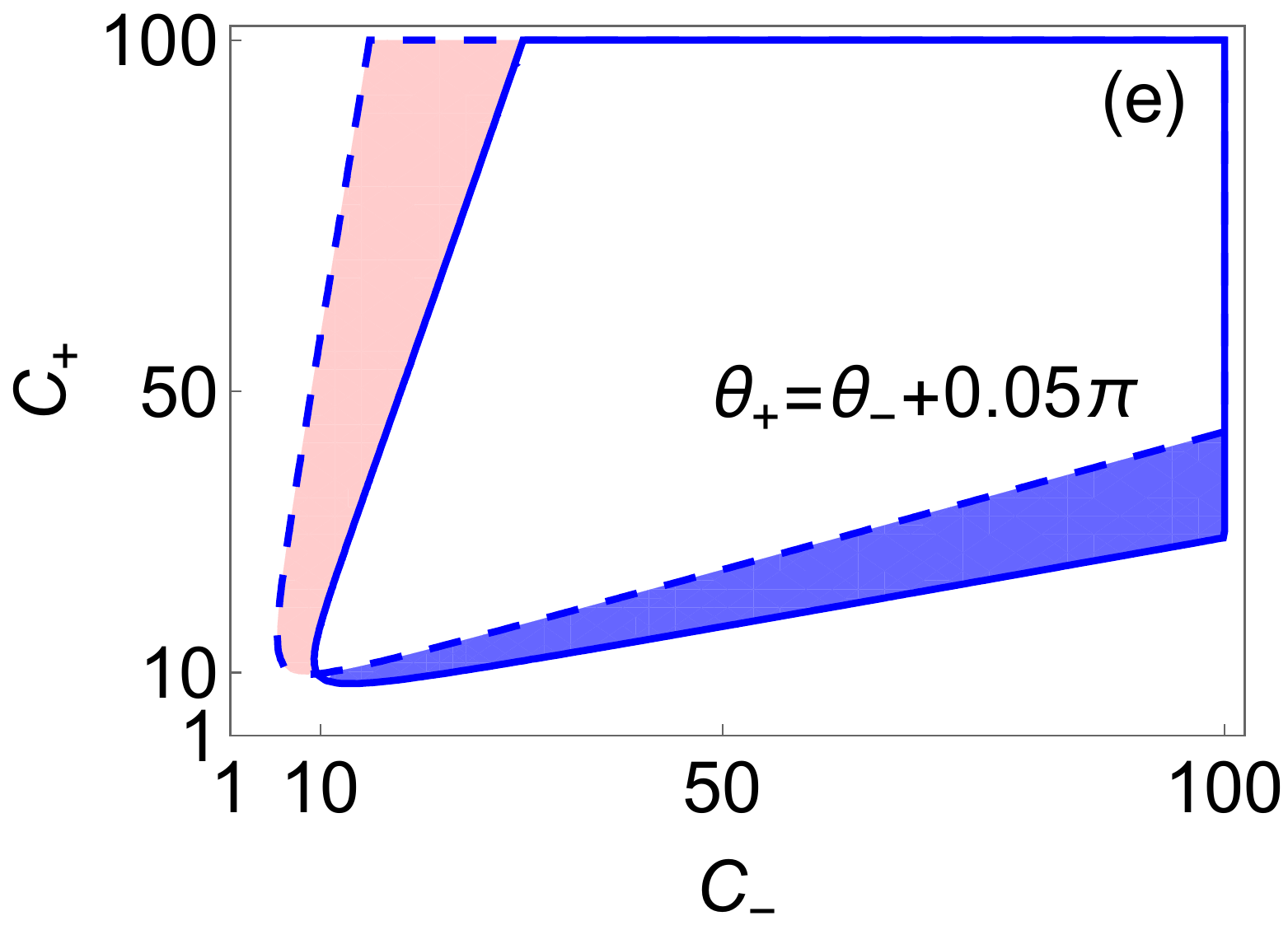}
\includegraphics[width=0.46\columnwidth]{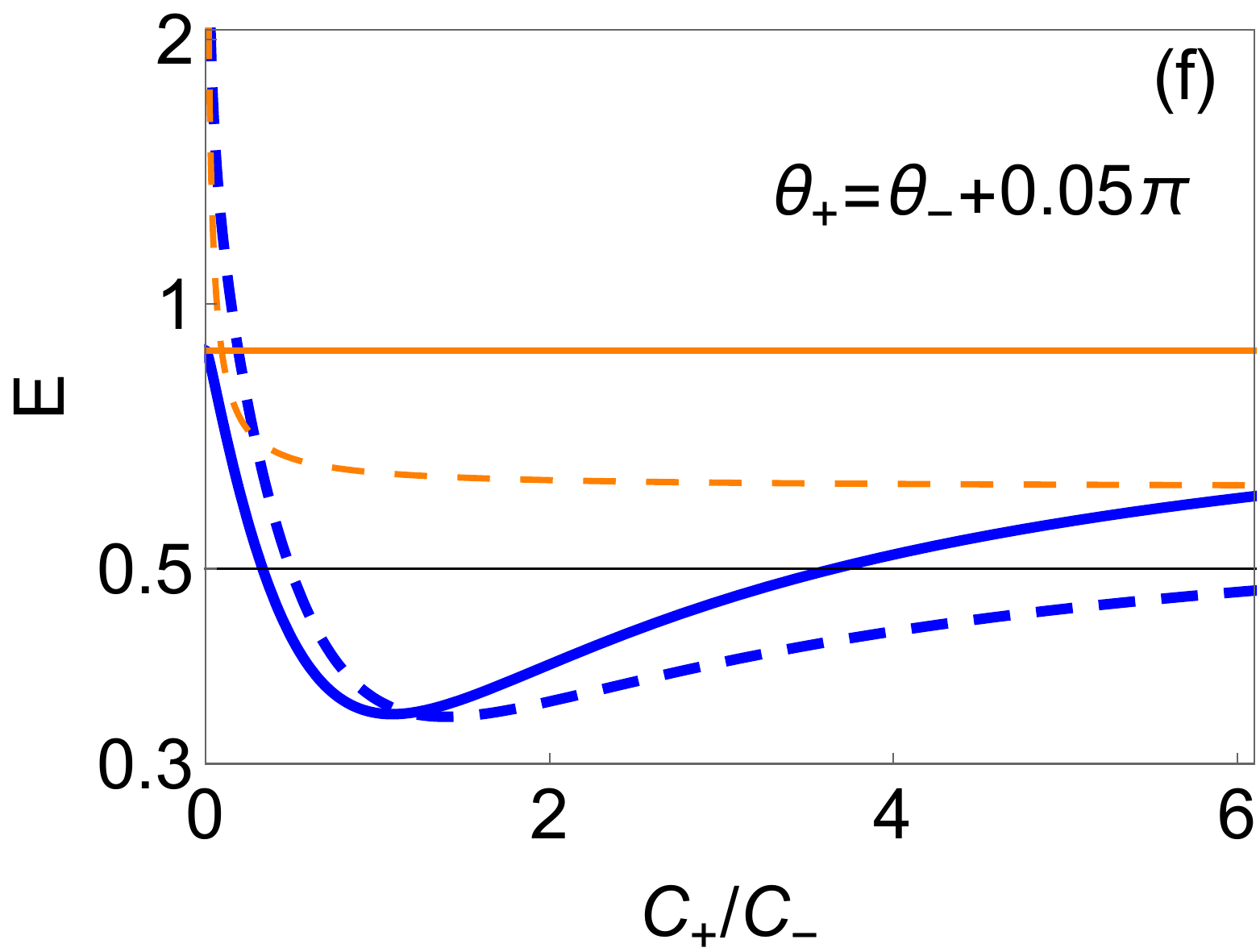}
}
\caption{(Left column) $E_{+ \mid -}$ (<0.5, in the region delineated by the dashed contour line) and $E_{- \mid +}$ (<0.5, solid contour line) as a function of the quantum cooperativities $C_{\pm}$. One-way steering from the first $(-)$ to the second $(+)$ subsystem exists in the light pink region, and one-way steering of the opposite direction exists in the dark blue region. (Right column) $E_{+ \mid -}$ (thick blue, dashed), $E_{- \mid +}$ (thick blue, solid) and the bare variances $\Delta^{2}\hat{X}_{+0}$ (thin orange, dashed), $\Delta^{2}\hat{X}_{-0}$ (thin orange, solid) of the two subsystems as a function of the ratio $C_{+}/C_{-}$ when assuming $C_{-}=50$. The remaining parameters are fixed to the following values:  $\theta_{-}=0.35\pi$, $\epsilon=0$, the intrinsic linewidth $\gamma_{0}=2\pi \times0.1\text{Hz}$, and the thermal occupation number $\bar{n}=10^{5}$. The thermal decoherence rate resulting from these thermal parameters is $\tilde{\gamma}_{0}\approx 2\pi \times 10 \text{kHz}$.}
\label{fig: ContourC}
\end{figure}

The contour plots in the left column of Fig.~\ref{fig: ContourC} show the regions in the $(C_{-},C_+)$ plane in which there exists two-way steering (white region), one-way steering for which $E_{+\mid-}<0.5<E_{-\mid+}$ (light pink region), and the opposite direction of one-way steering where $E_{-\mid+}<0.5<E_{+\mid-}$ (dark blue region). We remark that similar regions of steering are found when assuming instead that the thermal baths of the hybrid system are in the vacuum state, using the parameters values $\gamma_{0}=2\pi \times20\text{kHz}$ and $\bar{n}=0$, resulting in approximately the same decoherence rate $\tilde{\gamma}_{0}$ as considered in Fig.~\ref{fig: ContourC}.

The results indicate that tuning the quantum cooperativities of the subsystems plays different roles in controlling the directions of EPR steering under the three situations where $f=0$ ($\theta_+=\theta_-$), $f>0$ ($\theta_+=\theta_--0.05\pi$), and $f<0$ ($\theta_+=\theta_-+0.05\pi$). For the parameters considered here, when the two subsystems are driven optically only by the common vacuum field ($f=0$), two-way steering occurs for reasonably large quantum cooperativities, i.e., $C_{\pm}>10$, 
while the range of one-way steering is negligible, as shown in Fig.~\ref{fig: ContourC}(a). The reason is in this case that the asymmetry of steering builds up purely due to the difference of the bare variances of the two subsystems, i.e., $(\Delta\hat{X}_{\pm0})^2$ in Eqs.~(\ref{eq:E}), which are only determined by the difference of $C_{\pm}$ for the two subsystems. In the region where EPR steering appears, the asymmetry brought about by this difference is small. Distinct asymmetric steering appears in the presence of asymmetric noise interference in the two steering parameters as induced by the unidirectional coupling ($f\neq0$). 
As illustrated in Figs.~\ref{fig: ContourC}(c) and \ref{fig: ContourC}(e), both one-way steering in the direction $(- \Rightarrow +)$ (light pink region) and in the opposite direction $(+ \Rightarrow -)$ (dark blue region) can be achieved by suitably tuning the values of $C_{\pm}$ depending on the sign of $f$. 
%\textcolor{blue} {when $C_{-}\gtrsim 60$}. 
In addition, two-way steering (white region in the middle) still exists when  $0.40\apprle C_+/C_- \apprle1.7$ for $f>0$ and $0.45\apprle C_+/C_- \apprle3.2$ for $f<0$.

The precise values of the steering parameters $E_{+\mid-}$ (dashed) and $E_{-\mid+}$ (solid) are depicted in the right column of Fig.~\ref{fig: ContourC} as a function of $C_+$ for fixed $C_{-}=50$, 
corresponding to vertical cuts in the respective contour plots.
%To show this effect clearly, we plot the values of $E_{+\mid -}$ (blue, dashed) and $E_{-\mid+}$ (blue, solid) as a function of the ratio $C_{+}/C_{-}$ for a fixed $C_{-}$ (we choose $C_{-}=50$ as a feasible parameter in experiment~\cite{Krauter2011,Vasilakis2015}). 
Since in the decomposition of Eqs.~\eqref{eq:E}, the steering parameters $E_{+\mid -}$ and $E_{-\mid +}$ are composed of the bare variance and the noise reduction due to the correlation,
%the bare noise minus the amount of noise that can be decorrelated using the other system, 
we also plot the curves of $(\Delta\hat{X}_{+0})^2$ (thin orange, dashed) and $(\Delta\hat{X}_{-0})^2$ (thin orange, solid) as a function of the ratio $C_{+}/C_{-}$ to show this contribution to the steering parameters. % ($E_{+\mid -}$ and $E_{-\mid +}$). 
These curves reflect the dynamical cooling of the oscillators relative to their thermal equilibrium occupancy $\bar{n}_{\pm,0}$ as can be seen from Eq.~\eqref{eq:bare-var} when $\gamma_{\pm}>\gamma_{\pm,0}$. The bare variance $(\Delta\hat{X}_{+0})^2$ is decreased  (better dynamical cooling) with increasing  cooperativity $C_{+}$ for a fixed $\theta_{+} > \pi/4$, as shown in the right column of Fig.~\ref{fig: ContourC}, and its asymptote %of the bare noise $(\Delta\hat{X}_{+0})^2$ 
for large $C_{+}$ is the limit imposed by the finite degree of sideband asymmetry of the oscillator-light coupling as parametrized by $(\pi/4)\leq\theta_{+}\leq(\pi/2)$. In the fully asymmetric limit, $\theta_{+}=\pi/2$, the bare variance can be cooled to the ground state, i.e., $(\Delta\hat{X}_{+0})^2=1/2$, when $C_{+} \gg 1$ (referred to as the ``fully-resolved-sideband limit'' in the context of optomechanics).

For the performance of EPR steering, in the case of $f=0$ shown in Fig.~\ref{fig: ContourC}(b), the steerabilities in the two directions are approximately equal in the range of $C_{+}\gg1$. This can be understood by introducing the ratio of the steering parameters, which can be expressed using Eqs.~\eqref{eq:E} as
%To compare the performance of EPR steering in the two directions, we introduce the ratio of the two steering parameters to . From Eq.~\eqref{eq:E}, it can be derived as
\begin{equation}
\frac{E_{+\mid-}}{ E_{-\mid+}}=\frac{(\Delta\hat{X}_{+})^2}{(\Delta\hat{X}_{-})^2}=\frac{ (\Delta\hat{X}_{+0})^2+\sqrt{1-\epsilon}f\langle \hat{X}_{+},\hat{X}_{-}\rangle}{(\Delta\hat{X}_{-0})^2}.
\label{eq:ratio}
\end{equation}
When $f=0$ (i.e., $\theta_{\pm}=\theta$), the ratio of the steering parameters is equal to the ratio of the bare variances, which are both approximately equal to $-1/[2\cos(2\theta)]$ when optical broadening dominates $\gamma_{\pm} \approx \gamma_{\pm,\text{opt}}$ and cooperativities are large $C_{\pm}\gg1$. %since the degrees of sideband asymmetry are the same.}
However, for the cases of $f\neq0$ as shown in Figs.~\ref{fig: ContourC}(d) and \ref{fig: ContourC}(f), the steerabilities in the two directions become obviously asymmetric.
The reason for this asymmetry of steering can be seen from Eq.~\eqref{eq:ratio}, where the ratio of the steering parameters now depends not only on the different degrees of the dynamical cooling ($\theta_{+}\neq\theta_{-}$) but also on the asymmetric noise interference in the two steering parameters resulting from the unidirectional coupling ($f \neq0$).
 Specifically, when $f>0$ [Fig.~\ref{fig: ContourC}(d)],  $E_{+ \mid -}>E_{-\mid+}$ with the parameters studied here, making one-way steering possible only in the direction from the second $(+)$ to the first $(-)$ subsystem.
In contrast, for the case of $f<0$ [Fig.~\ref{fig: ContourC}(f)], 
we find four regions of steering with increasing $C_{+}/C_{-}$: no steering, one-way steering from the second $(+)$ to the first $(-)$, two-way steering, and again one-way steering in the opposite direction, indicating that the asymmetry of steering can be controlled more flexibly via tuning the quantum cooperativities $C_{\pm}$ when $f<0$.

\subsection{Tuning interaction angles}

\begin{figure}[!h]
\centering{
\includegraphics[width=0.48\columnwidth]{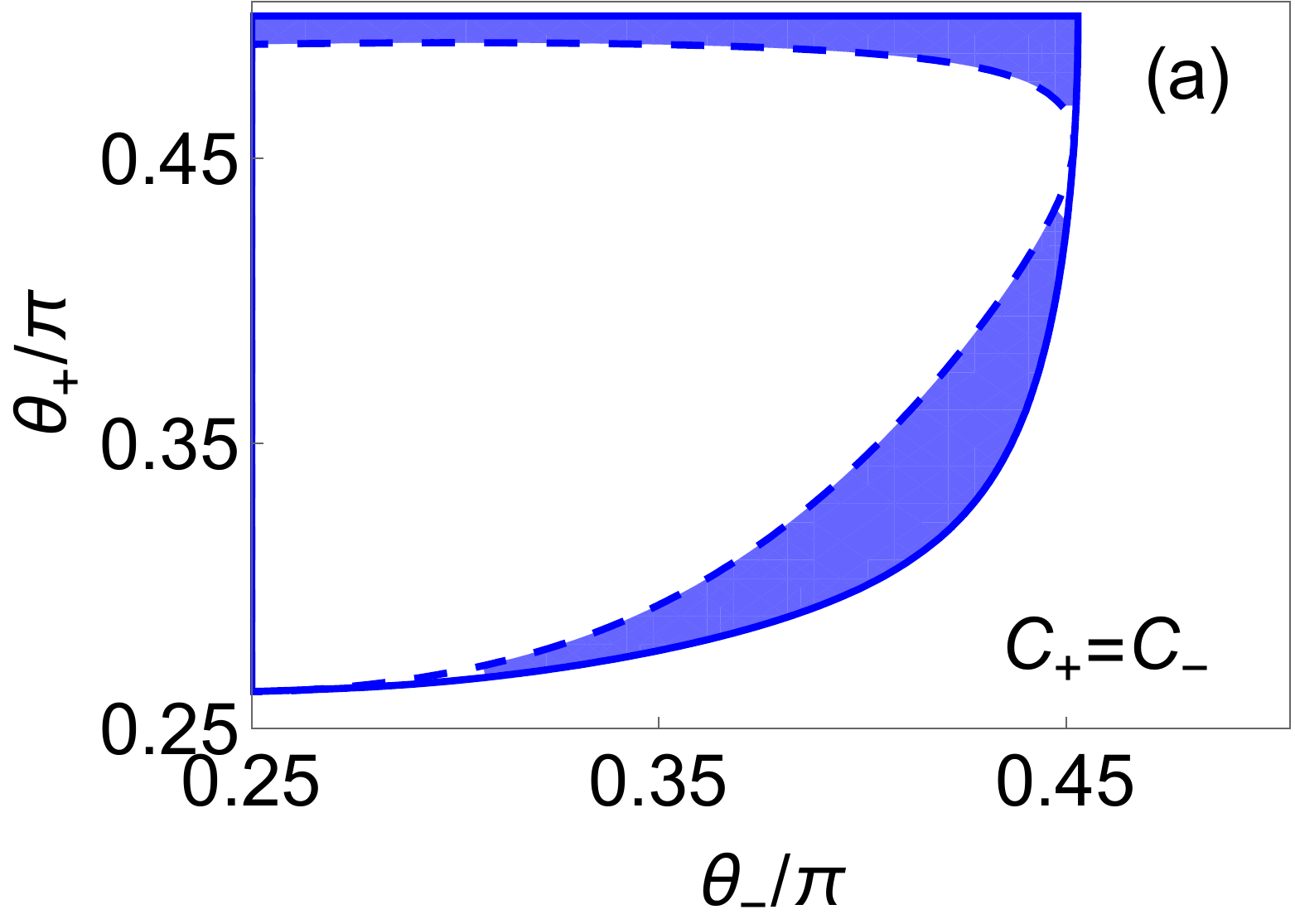}
\includegraphics[width=0.48\columnwidth]{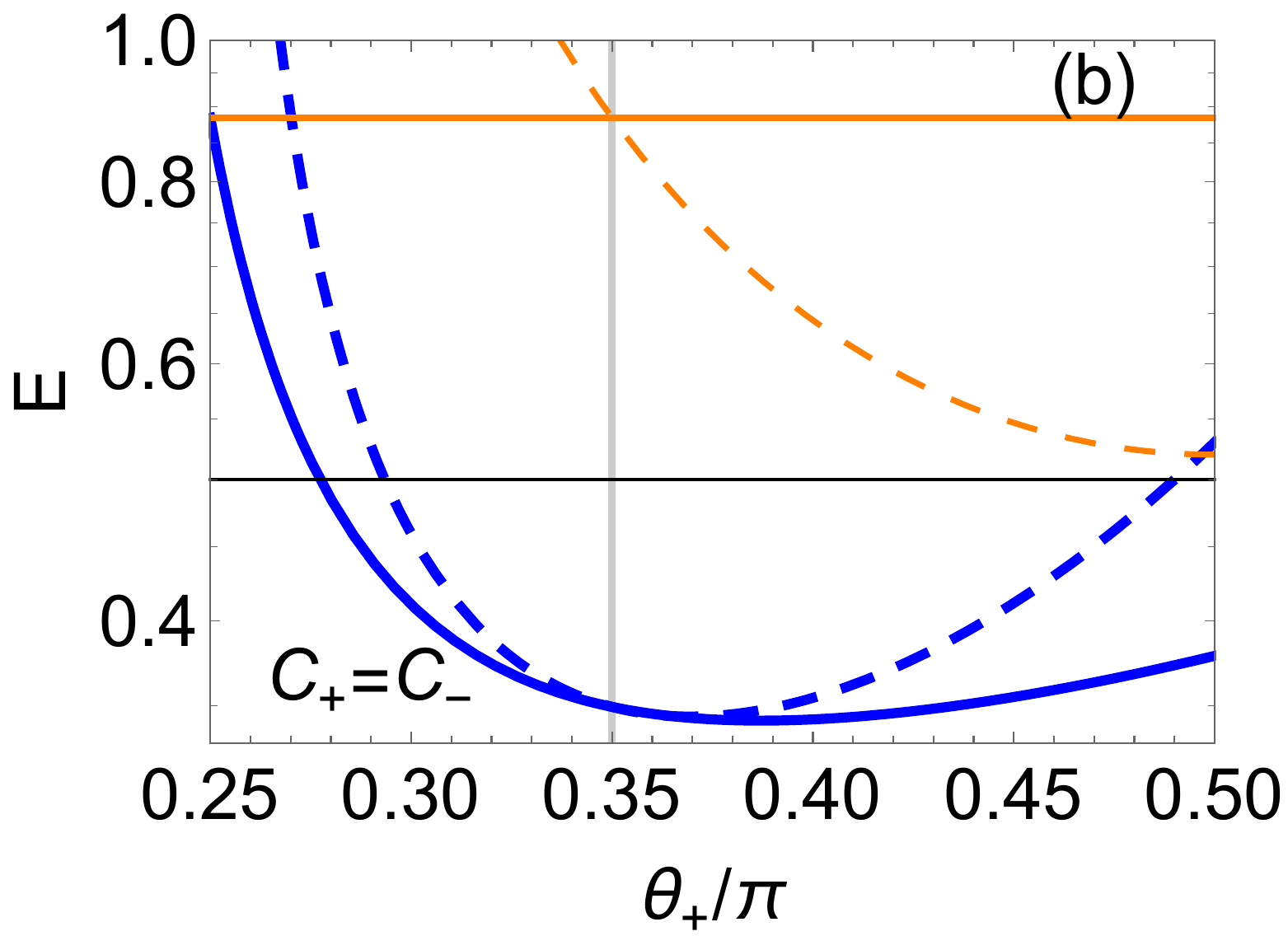}
\includegraphics[width=0.48\columnwidth]{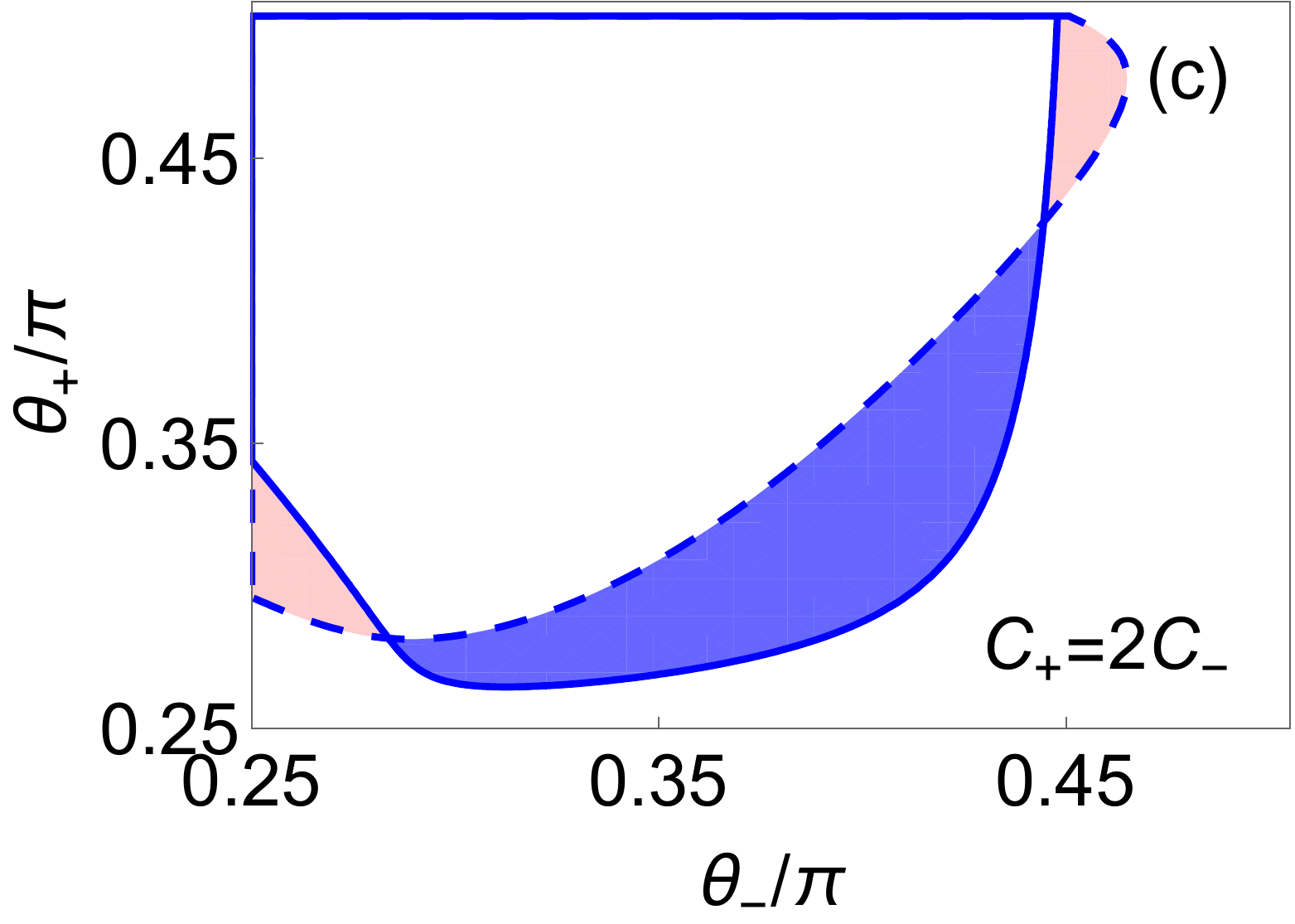}
\includegraphics[width=0.48\columnwidth]{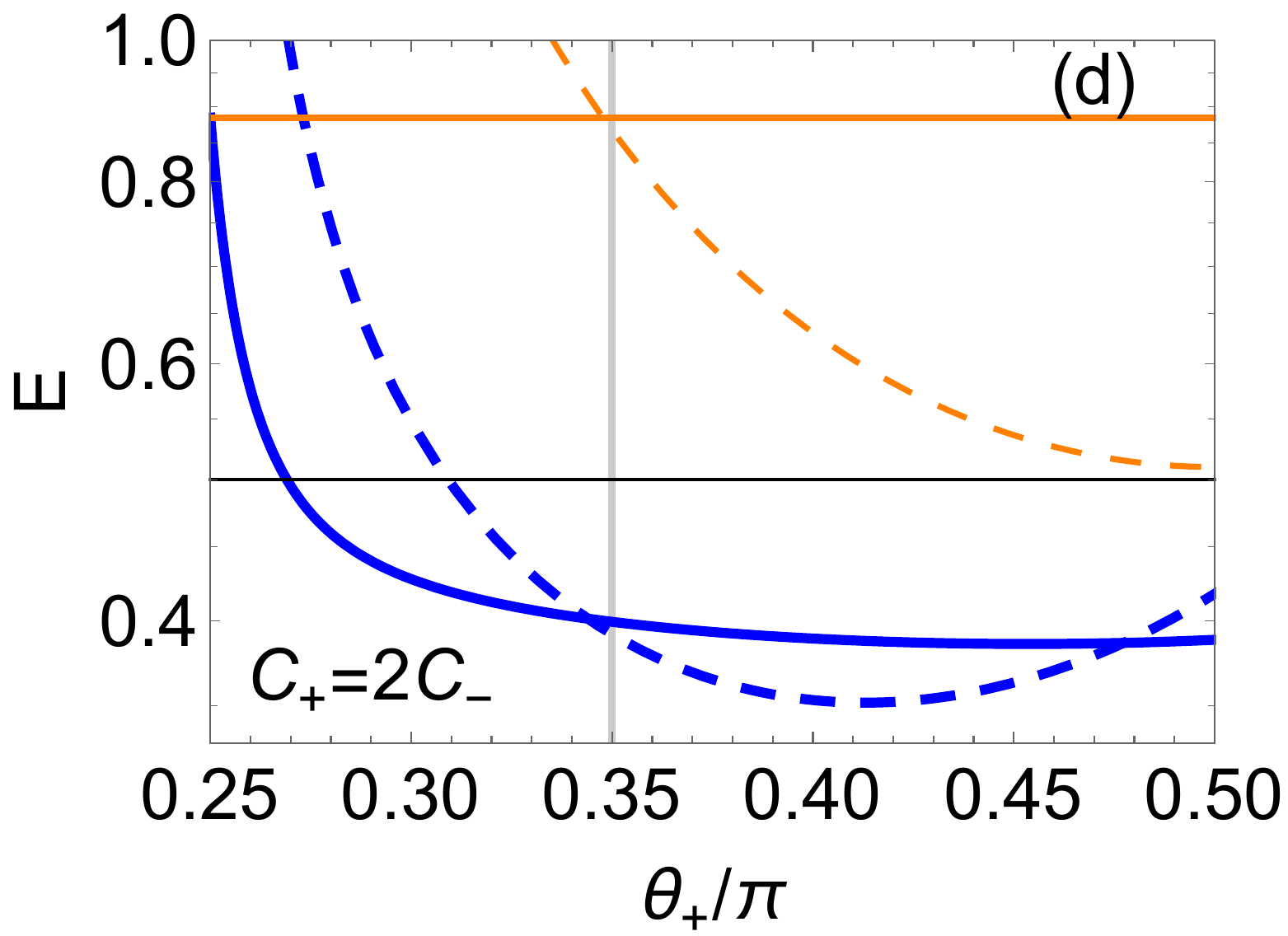}
}
\caption{(Left column) $E_{+ \mid -}$ (<0.5, in the region delineated by the dashed contour) and $E_{- \mid +}$ (<0.5, solid contour) as a function of interaction angles $\theta_{\pm}$ for (a) $C_{+}=C_{-}$ and (c) $C_{+}=2C_{-}$. One-way steering from the first $(-)$ to the second $(+)$ subsystem exists in the light pink region, and one-way steering of the opposite direction exists in the dark blue region.
(Right column) $E_{+ \mid -}$ (thick blue, dashed), $E_{- \mid +}$ (thick blue, solid) and the bare variances in the two subsystems $\Delta^{2}\hat{X}_{+0}$ (thin orange, dashed), $\Delta^{2}\hat{X}_{-0}$ (thin orange, solid) as a function of  $\theta_{+}$ when assuming $\theta_{-}=0.35\pi$; again, we consider the two cases (b) $C_{+}=C_{-}$ and (d) $C_{+}=2C_{-}$. The other parameter values used are  $C_{-}=50$, transmission loss $\epsilon=0$, intrinsic linewidth $\gamma_{0}=2\pi \times0.1\text{Hz}$, and thermal occupation number $\bar{n}=10^{5}$, resulting in the decoherence rate $\tilde{\gamma}_{0}\approx 2\pi \times 10 \text{kHz}$.}
\label{fig: ContourT}
\end{figure}

We now show the behavior of EPR steering as a function of interaction angles $\theta_{\pm}$ for fixed values of $C_{\pm}=50$ and $2C_{-}=C_{+}=100$. As illustrated in Fig.~\ref{fig: ContourT}, for the parameters considered here, one can find suitable ranges of $\theta_{\pm}$ to realize both one-way and two-way steering. To be specific, when $C_{\pm}=50$, two-way steering (white region) and one-way steering in the direction $ + \Rightarrow -$ (dark blue region) can be achieved by tuning  $\theta_{\pm}$, as shown in Fig.~\ref{fig: ContourT}(a). However, when $C_{+} =2 C_{-}=100$ [Fig.~\ref{fig: ContourT}(c)], 
one-way steering in both the direction $ + \Rightarrow -$ (dark blue region) and $ - \Rightarrow +$ (light pink region) become possible by tuning $\theta_{+}<\theta_{-}$ and $\theta_{+}>\theta_{-}$, respectively, and these regions meet near the crossing points $\theta_{+} \approx \theta_{-}$ (i.e., $f\approx 0$). Note that similar regions of EPR steering in the $(\theta_{-},\theta_{+})$ plane are found if instead assuming the subsystems to be coupled to vacuum thermal baths, i.e., if the thermal parameters are assumed to be $\gamma_{0}=2\pi \times20\text{kHz}$ and $\bar{n}=0$, thereby matching the decoherence rate $\tilde{\gamma}_{0}$ studied in Fig.~\ref{fig: ContourT} (as was similarly found for the steering regions in terms of $C_{\pm}$ above [left column of Fig.~\ref{fig: ContourC}]).

Next, in Figs.~\ref{fig: ContourT}(b) and~\ref{fig: ContourT}(d), we show the bare variances $(\Delta\hat{X}_{\pm0})^2$ and the steerabilities $E_{+ \mid -}$ and $E_{- \mid +}$ as functions of the interaction angle $\theta_{+}$ while fixing $\theta_{-}=0.35\pi$. Since higher degree of the sideband asymmetry (larger $\theta_{+}$) corresponds to better dynamical cooling for fixed $C_{+}$ [Eqs.~\eqref{eq:ss_EPRvar}], $(\Delta\hat{X}_{+0})^2$ is decreased with increasing $\theta_{+}$ as shown in Figs.~\ref{fig: ContourT}(b) and~\ref{fig: ContourT}(d). 
Meanwhile, % for the steering parameters ($E_{+ \mid -}, E_{- \mid +}$), 
due to the combined effects of the different degrees of the dynamical cooling and the induced asymmetric noise interference %in the two steering parameters 
shown in Eqs.~\eqref{eq:E}, the behavior of $E_{+ \mid -}$ and $E_{- \mid +}$ with increasing $\theta_{+}$ is generally not monotonic. Optimal choices of $\theta_{+}$ for maximizing the steerabilities exist in the region of $f<0$, which we will explore in the next section.\\

\section{Steady-state steering optimization} \label{sec: Optimization}

Since EPR steering plays a key role in the implementation of 1sDI quantum tasks, in this section we apply and optimize our scheme as a steering resource for such applications. The particular scenario we will consider is that in which an untrusted client node wants to connect to a trusted server node, hence requiring steering from the former to the latter system in order to enable 1sDI tasks.
In order to restrict the parameter space of our optimization, we will make assumptions about the accessible range of parameters for the two subsystems. To this end, we imagine a quantum network in which a number of relatively cheap client nodes are connected to a relatively expensive central server node.
It seems reasonable to assume that a ``cheap'' client node has limited ranges of interaction strength and type with the optical field (parametrized by $C$ and $\theta$), as compared to the ``expensive'' server node, and we therefore take these client parameters to be the constrained parameters in the steering optimization below (while assuming the server parameters to be freely tunable). Under these circumstances, we find that the optimal cascade ordering is to let the client be the first subsystem and the server the second subsystem, corresponding to labels $(-)$ and $(+)$, respectively, in preceding sections; henceforth, we focus on this case. While the steering from client to server node is the quantity of primary interest in the present section, as motivated above, we will for comparison also optimize and plot the steering in the opposite direction.

\begin{figure}[!h]
\centering{\includegraphics[width=0.49\columnwidth]{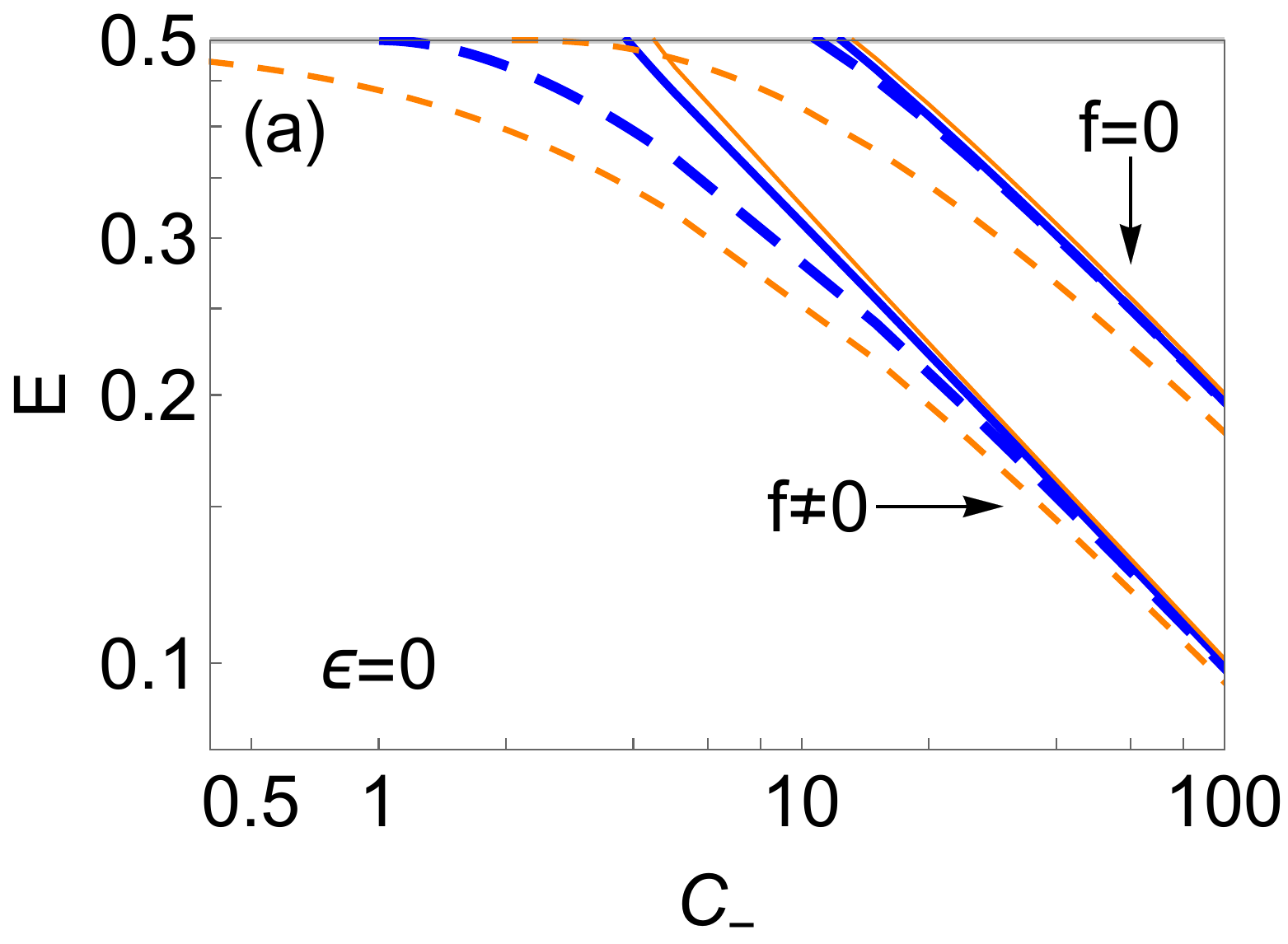}
\includegraphics[width=0.49\columnwidth]{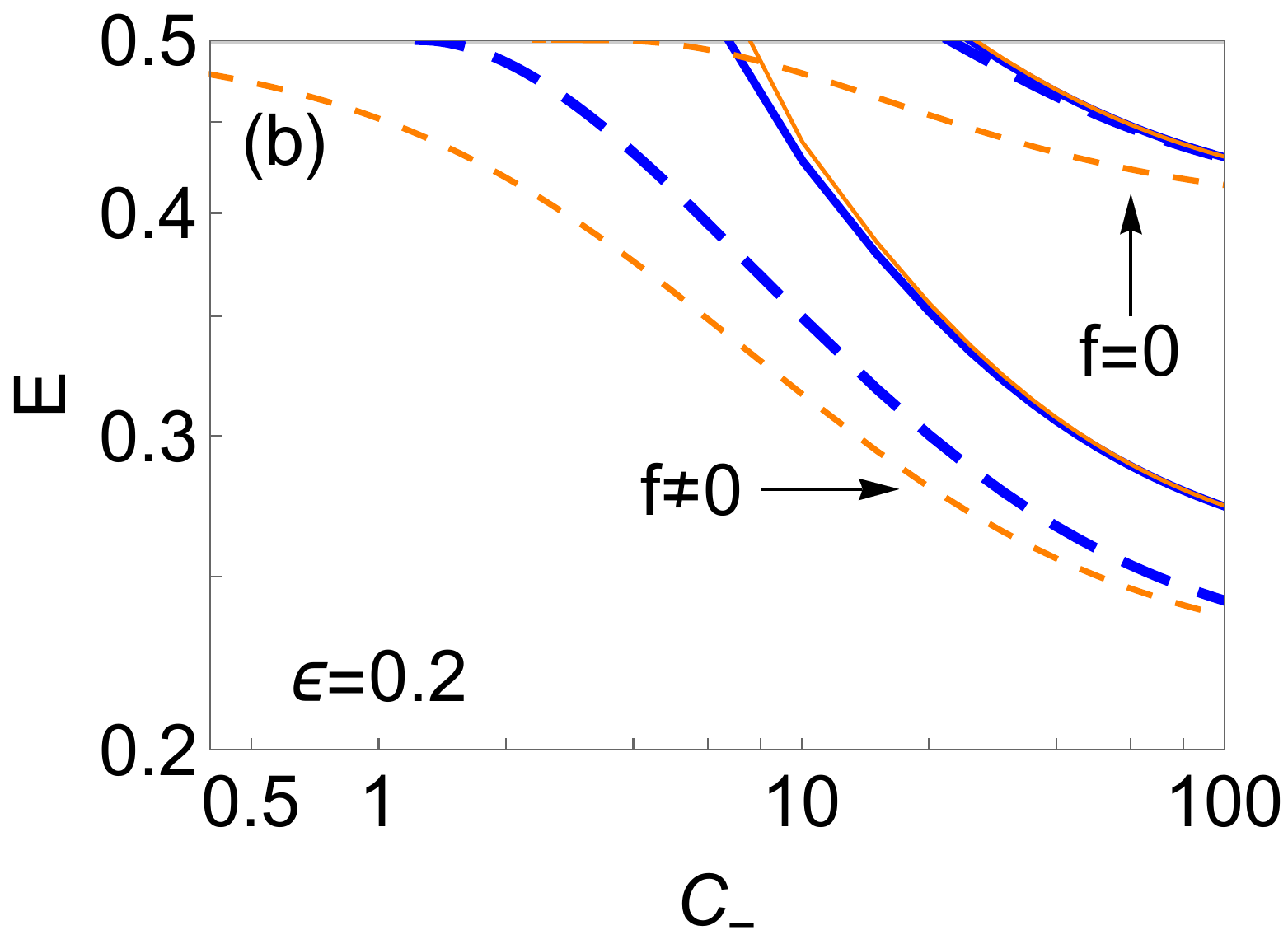}}
\caption{$E_{+ \mid -}$ (dashed) and $E_{- \mid +}$ (solid) as a function of $C_{-}$ with optimal interaction types $\theta_{\pm}$ and quantum cooperativity $C_{+}$ when $f=0$ and $f\neq 0$, in the cases (a) $\epsilon=0$ and (b) $\epsilon=0.2$, i.e., without and with transmission losses. The two sets of thermal parameters are (thick blue curves) intrinsic linewidth $\gamma_{\pm,0}=2\pi \times0.1\text{Hz}$, thermal occupation number $\bar{n}_{\pm}=10^{5}$; (thin orange curves) $\gamma_{-,0}=2\pi \times0.1\text{Hz}$ , $\gamma_{+,0}=2\pi \times20\text{kHz}$, $\bar{n}_{-}=10^{5}$ and $\bar{n}_{+}=0$.}
%when $f=0$ (red) and $f \neq 0$ (blue), in the cases (a) $\epsilon=0$ and (b) $\epsilon=0.2$, i.e., without and with transmission losses. The remaining (thin) curves represent the scenario of the asymmetric thermal parameters with intrinsic linewidths $\gamma_{-,0}=2\pi \times0.1\text{Hz}$ and $\gamma_{+,0}=2\pi \times20\text{kHz}$, and thermal occupation numbers $\bar{n}_{-}=10^{5}$ and $\bar{n}_{+}=0$, when $f=0$ (black) and $f \neq 0$ (orange).}
\label{fig: opt_E}
\end{figure}

Given the above considerations, we proceed by separately optimizing the steerability in each direction, $E_{+ \mid -}$ and $E_{- \mid +}$, for a given value of the client cooperativity, $C_{-}$, i.e., evaluating the remaining parameters $C_{+}$ and $\theta_{\pm}$ at their optimal values (these optimal parameter values are plotted in Appendix~\ref{app:optimal-params}). 
The resulting minimal values of $E_{\pm \mid \mp}$ 
%The numerical minimization of the EPR-steering parameters 
presented in Fig.~\ref{fig: opt_E} show that optimizing the unidirectional coupling of the first $(-)$ to the second $(+)$ subsystem induced by asymmetric coupling types $f \neq 0$ (presented by the blue and orange curves) can significantly enhance the steerability in both directions (indicated by the smaller values of $E_{+\mid-}$ and $E_{-\mid+}$), as compared with the case of symmetric coupling, $f=0$ (i.e., forcing $\theta_{\pm}=\theta$, presented by the red and black curves).
The corresponding optimal asymmetric interaction angles (resulting in the blue and orange curves) show that in the regime where $\bar{n}_{-}\gg1$, the optimal interaction type between the first subsystem $(-)$ and the light field is QND interaction ($\theta_-\approx\pi/4$) for all $C_{-}\gtrsim 1$; this corresponds to no sideband asymmetry ($\Gamma_{-B}=\Gamma_{-P}$), a condition which will typically be easy to fulfill and hence is compatible with our assumption of the ``cheap'' client system having limited tunability of its interaction type, $\theta_{-}$ (this is true for, e.g., unresolved-sideband optomechanical systems and free-space spin ensembles). 
Meanwhile, for the second subsystem $(+)$, more cooling (beam-splitter-type) than heating (parametric-down-conversion-type), i.e., $\theta_{+}>\pi/4 \Leftrightarrow \Gamma_{+B}>\Gamma_{+P}$, is required. Taken together, these optimal values for $\theta_{\pm}$ indicate that choosing $f<0$ is beneficial for maximizing the steerabilities in our hybrid system. % (plots of the optimal values of $\theta_{\pm}$ as well as the required server cooperativity $C_{+}$ are given in Appendix A).
This result is consistent with that presented in Ref.~\cite{Huang2018}, i.e., the additional interference effect allowed by the optimal interaction angles can be used to achieve the maximum noise cancellation in the EPR variables, and thus enhance the correlation between the two subsystems.

Having established the advantage of the unidirectional coupling corresponding to $f<0$, we now turn to the asymmetry between the optimal steering achievable in the two directions.
%To focus on the enhancement of the optimal steerabilities in a particular direction,
As seen in Fig.~\ref{fig: opt_E}, we find the optimal client-to-server steerability $E_{+ \mid -}$ to significantly outperform the opposite direction $E_{- \mid +}$ for moderate client cooperativities $C_{-}\lesssim 10$ in the absence of transmission losses~[Fig.~\ref{fig: opt_E}(a)] and for all $C_{-}$ in the presence of moderate losses~[Fig.~\ref{fig: opt_E}(b)]; 
%their difference $E_{- \mid +}-E_{+ \mid -}>0$ for a small $C_{-}$ to be significantly enhanced by choosing asymmetric interaction types. 
the situation is qualitatively the same whether the server system's thermal bath is assumed to be hot (blue curves) or vacuum (orange curves), while keeping the decoherence rate $\tilde{\gamma}_{+,0}$ fixed, although the latter scenario is seen to be preferable.
%Moreover, the enhanced difference independent of the thermal parameters of the second subsystem $(+)$, which can be either thermally hot or near the ground state, as its cooperativity ($C_{+}$) and sideband asymmetry ($\theta_{+}$) can be tuned to implement the optimal dynamical cooling (ground-state cooling) of the thermal noise.  
%The asymmetry in the optimal steerabilities ($E_{+ \mid -}, E_{- \mid +}$) leads to the enhanced one-way steering from the first ($-$) to the second $(+)$ subsystem, 
It follows from these results that our scheme is especially advantageous in the technological scenario envisaged here, i.e., producing the steering required for 1sDI quantum communications between a poorly tunable untrusted client node and a highly tunable trusted server node.

Considering the effect of transmission losses $\epsilon >0$ [Fig.~\ref{fig: opt_E}(b)], 
%the steerabilities are reduced but does not result in additional difference of the optimized steering degrees in the two directions when $f=0$; 
the optimal steerabilities are degraded, but to different degrees in the two directions for $f\neq 0$, even in the limit of large $C_{-}$. 
This is to be expected from Eqs.~\eqref{eq:E}, where $\epsilon$ enters asymmetrically in $E_{+\mid-}$ and $E_{-\mid+}$, and hence will affect the optimal steering in the two directions differently when $f\neq0$.
%Due to the cascaded setting and the presence of the transmission loss $\epsilon$, part of the unidirectional light field output from the first subsystem is replaced by the extra vacuum noise such that the driving from the first to the second subsystem is reduced. 
To be specific, besides the reduction of the covariance term $|\langle \hat{X}_{+},\hat{X}_{-}\rangle|$, %which enters symmetrically in $E_{+\mid-}$ and $E_{-\mid+}$~[Eq.~\eqref{eq:EPRMSmin} and~\eqref{eq:EPRSMmin}], 
non-zero $\epsilon$ will also affect the additional noise interference, and thus change $(\Delta\hat{X}_{+})^2$, while it has no influence on $(\Delta\hat{X}_{-})^2$. In Fig.~\ref{fig: opt_E}b, this asymmetry of nonzero $\epsilon$ is seen to increase the separation $E_{+\mid-} < E_{-\mid+}$ between the optimal steerabilities. 
In fact, considering the transmission loss $\epsilon$ as an effective external loss added to the second ($+$) oscillator, this result is consistent with that of earlier papers~\cite{Reid2009,Handchen2012,Rosales-Zarate2015}, i.e., the decoherence of EPR steering is substantial when the loss is on the steering object, but much less sensitive to the loss on the steered object. The loss is added to the second ($+$) oscillator which acts as the steering object in $E_{-\mid+}$, but as the steered party in $E_{+\mid-}$, such that the loss reduces the correlation in the direction $+ \Rightarrow -$ more than the other direction, i.e., the value of $E_{-\mid+}$ is higher than $E_{+\mid-}$. This asymmetric effect of loss with respect to the two subsystems is another consequence of the inherently asymmetric nature of EPR steering, which does not apply to entanglement.

\begin{figure}[!h]
\includegraphics[width=0.6\columnwidth]{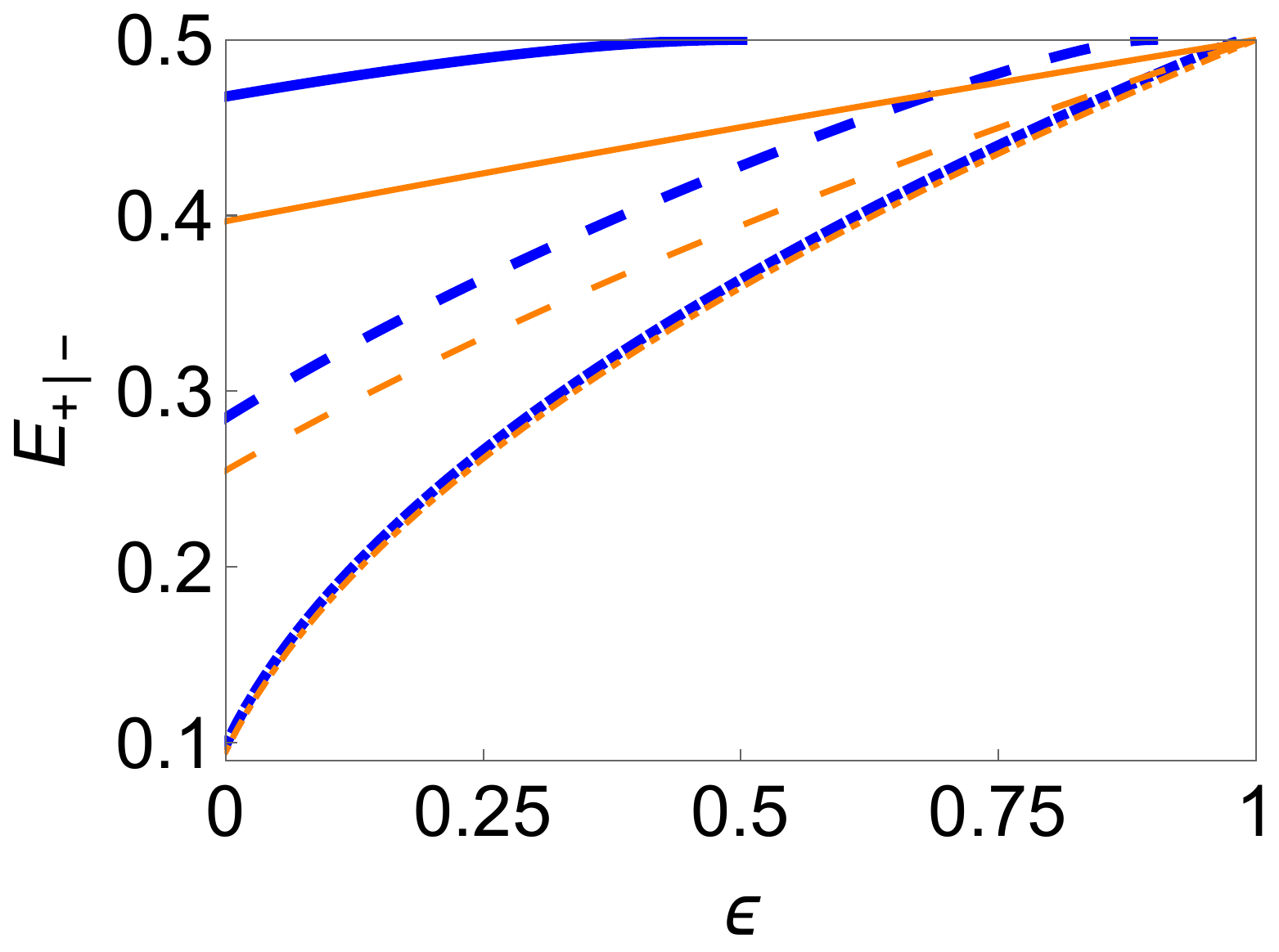}
\caption{$E_{+ \mid -}$ as a function of the transmission loss $\epsilon$ evaluated for optimized $\theta_{\pm}$ and $C_{+}$ when $C_{-}=2$ (solid), 10 (dashed), and 100 (dot-dashed). The two sets of fixed thermal parameters are (thick blue curves) intrinsic linewidth $\gamma_{\pm,0}=2\pi \times0.1\text{Hz}$ and thermal occupation number $\bar{n}_{\pm}=10^{5}$; (thin orange curves) $\gamma_{-,0}=2\pi \times0.1\text{Hz}$, $\gamma_{+,0}=2\pi \times20\text{kHz}$, $\bar{n}_{-}=10^{5}$, and $\bar{n}_{+}=0$.}
%Optimal interaction types $\theta_{\pm}$ and ratio of cooperativities $C_{+}/C_{-}$ that minimize $E_{+ \mid -}$ (dashed) and $E_{- \mid +}$ (solid) as a function of $C_{-}$ when $R=0$ (blue) and $R \neq 0$ (orange).  The fixed parameters are assumed: intrinsic linewidth $\gamma_{0}=2\pi \times0.1\text{Hz}$, thermal occupation number $\bar{n}=10^{5}$.
\label{fig:loss}
\end{figure}

Since transmission losses are a crucial factor in any quantum network architecture, %implementation of distributed quantum tasks. 
in Fig.~\ref{fig:loss} we study further the robustness of our scheme in this regard, focusing on the optimized client-to-server steerability $E_{+ \mid -}$. %we illustrate its effect on steering from the untrusted client node ($-$) to the trusted server node ($+$). 
The tolerance to transmission losses $\epsilon$ is seen to depend on both the client cooperativity $C_{-}$ and the thermal parameters of the server, i.e., the intrinsic linewidth $\gamma_{+,0}$ and the thermal occupation $\bar{n}_{+}$.
In the case of large cooperativity, $C_{-}=100$, $E_{+\mid-}$ is almost equal for the two sets of thermal parameters considered in Fig.~\ref{fig:loss}. However, when $C_{-}$ is small or moderate, the case that assumes vacuum-state thermal parameters for the server $(+)$ subsystem  (orange curves) can tolerate more loss than the case of hot thermal parameters (blue curves) while achieving the same $E_{+\mid-}$. The maximum loss tolerance for the case of the hot thermal parameters for the server $(+)$ subsystem is $\epsilon<0.9$ when $C_{-}=10$ and $\epsilon<0.5$ when $C_{-}=2$. Note that $E_{+\mid-}<0.5$ can be achieved for arbitrary $\epsilon<1$ when assuming vacuum-state thermal parameters for the server $(+)$ subsystem.\\

\section{Optimality and implementation} \label{sec: discussion}

As a final but essential part of our analysis, we address the question of optimality for our unconditional scheme. In this regard, a natural question is whether our scheme wastes information by discarding the propagating light field after it has interacting with both subsystems (Fig.~\ref{fig:set up}). 
To investigate this, we have recalculated the optimal steerabilities (considered in Fig.~\ref{fig: opt_E}) that can be achieved when additionally measuring the joint output light field and optimally combining this measurement record with the verification measurements of the individual subsystems. %(see details in Appendix B). 
Mathematically, this is achieved by solving the corresponding stochastic master equation, yielding covariance matrix elements generalizing Eqs.~\eqref{eq:ss_EPRvar} to incorporate the information gained from the joint output field. The details of this procedure for the hybrid system under consideration here are given in the Supplementary Material of Ref.~\cite{Huang2018} and hence will not be reproduced here.
Evaluating this conditional scheme and optimizing the performance under the same constraints as in Fig.~\ref{fig: opt_E} (in particular, demanding dynamical stability), we find that the improvement in the steerabilities due to the added measurement of the joint output field is negligible compared to our unconditional scheme (plots are presented in Appendix~\ref{app:Cond-comparison}); this is similar to what has been found for entanglement generation in this hybrid system~\cite{Huang2018}.
One can interpret this outcome as being due to the second subsystem in the cascade acting as a coherent measurement device, rendering additional measurements on the joint field superfluous. This establishes the advantage of our scheme that the optimal steerabilities %in such hybrid systems 
can be obtained unconditionally (however, further improvements are likely possible by adding active feedback stabilization~\cite{Vasilyev2013}).

Having completed the analysis of our generic scheme, we now mention a couple of candidate physical systems for implementing it: collective spin ensembles and mechanical systems.
For both kinds of systems, the interaction strength and type with the light field (i.e., $\Gamma$ and $\theta$) can be efficiently manipulated by embedding them in a suitable optical cavity~\cite{Aspelmeyer2014,Vasilakis2015, Kohler2017}. 
Current state-of-the-art optomechanical experiments have demonstrated $C \gtrsim 30$~\cite{Mason2019}, although in the unresolved-sideband regime. If an optomechanical system is used to implement the second system in the cascade (Fig.~\ref{fig:set up}) then, generally, a rather sideband-resolved cavity is required in order to attain the optimal performance discussed in Sec.~\ref{sec: Optimization}.
Note that in the context of optomechanics, the quantum cooperativity $C\propto (g_0/\kappa)^2 |\alpha|^{2}\kappa/\gamma_{M0}$, where $g_{0}$ is the single-photon coupling rate and $\kappa$ is the cavity decay rate.  Hence, large values $C>1$ can be achieved by increasing the intracavity drive field $\alpha$ or decreasing the mechanical intrinsic dissipation $\gamma_{M0}$ while keeping the ratio $g_{0}/\kappa<1$, so that the system remains in the linearized optomechanical regime~\cite{Aspelmeyer2014}. %so that the approximation of the cavity field fluctuations being much smaller than the drive-induced amplitude $\alpha \gg \hat{a}_{\text{cav}}$ remains valid.}
Concerning spin ensembles, cooperativities $C\gtrsim 1$ are feasible~\cite{Krauter2011,Moller2017} and larger values are achievable by means of cavity enhancement~\cite{Vasilakis2015,Kohler2017}. However, such enhancement typically comes at the cost of increased optical losses (effectively larger $\epsilon$), prompting an experimental trade-off between large $C$ and small $\epsilon$.  
%and in either case $C \gtrsim 30$ is realistic in current state-of-the-art experiments~\cite{Vasilakis2015,Mason2018}. 

The candidate systems discussed here have inspired the choices of thermal parameter sets used in the plots above, as we will now describe.
The effective thermal occupation of a collective spin oscillator is due to imperfect optical pumping, which is typically equivalent to a near-vacuum bath, $\bar{n}\lesssim 1$. Additionally, the power broadening induced by the spontaneous emission can result in large intrinsic linewidths of such spin oscillators~\cite{Moller2017}, $\gamma_0/(2\pi) \gtrsim 1\,\text{kHz}$. 
%The decoherence of spin systems is typically composed of spin collisions and imperfect polarization, which is equivalent to a coupling with a (near-)vacuum bath with some large intrinsic damping rate (due to optical power broadening)~\cite{Hammerer2010}. 
For the mechanical oscillators, the thermal occupation is determined by the environment temperature and the resonance frequency, leading in most instances to $\bar{n}\gg 1$. However, small intrinsic damping rates corresponding to quality factors $Q>10^{6}$ are routinely achieved~\cite{Aspelmeyer2014} and the state of the art is $Q\gtrsim 10^9$~\cite{Mason2019}. For mechanical oscillators with $\Omega/(2\pi)\sim 1\,\text{MHz}$, this amounts to intrinsic linewidths in the range $\gamma_0/(2\pi)\sim 1\,\text{mHz---} 1\,\text{Hz}$.

\section{Conclusion}\label{sec: conclusion}

In conclusion, we have analyzed how to unconditionally generate tunable symmetric and asymmetric steady-state EPR steering in macroscopic hybrid systems composed of two thermal oscillators with masses of opposite signs coupled to a unidirectional light field. We have developed an approach to controlling the asymmetry of EPR steering by tuning the directional coupling of the first to the second subsystem resulting from the asymmetric interaction types between the subsystems and the light field. Instead of the methods of adding extra asymmetric amounts of noises or losses to each subsystem, our scheme provides an active method for the control of asymmetry of steering within the apparatus itself. Investigating possible applications, we find that our scheme can be used to engineer enhanced steering from an untrusted node of limited tunability to a trusted, highly tunable node, thus allowing for 1sDI quantum information tasks in macroscopic hybrid quantum networks.\\

\begin{acknowledgements}
The authors acknowledge E.\,S.\,Polzik, R.\,A.\,Thomas, and M.\,Fadel for helpful discussions. This work was supported by the National Key R$\&$D Program of China (Grants No.\,2018YFB1107200 and No.\,2016YFA0301302) and the National Natural Science Foundation of China (Grants No.\,11622428, No.\,61675007, and No.\,61475006), and the Key R$\&$D Program
of Guangzhou Province (Grant No.\,2018B030329001). E.\,Z.~is supported by the Carlsberg Foundation. Q.\,H.~thanks the Beijing Computational Science Research Center for the hospitality.\\
\end{acknowledgements}

\begin{appendix}

\section{Optimal parameters for EPR steering in a particular direction}\label{app:optimal-params}

In this section, we present the details of the numerical minimization of $E_{+\mid-}$ and $E_{-\mid+}$, as shown in Fig.~\ref{fig: opt_E} in the main text.
In particular, we plot the optimal values of the interaction angles $\theta_{\pm}$ and the cooperativity ratio $C_{+}/C_{-}$ required to realize the minimal $E_{\pm\mid\mp}$.

\begin{figure}[!h]
\centering{\includegraphics[width=0.49\columnwidth]{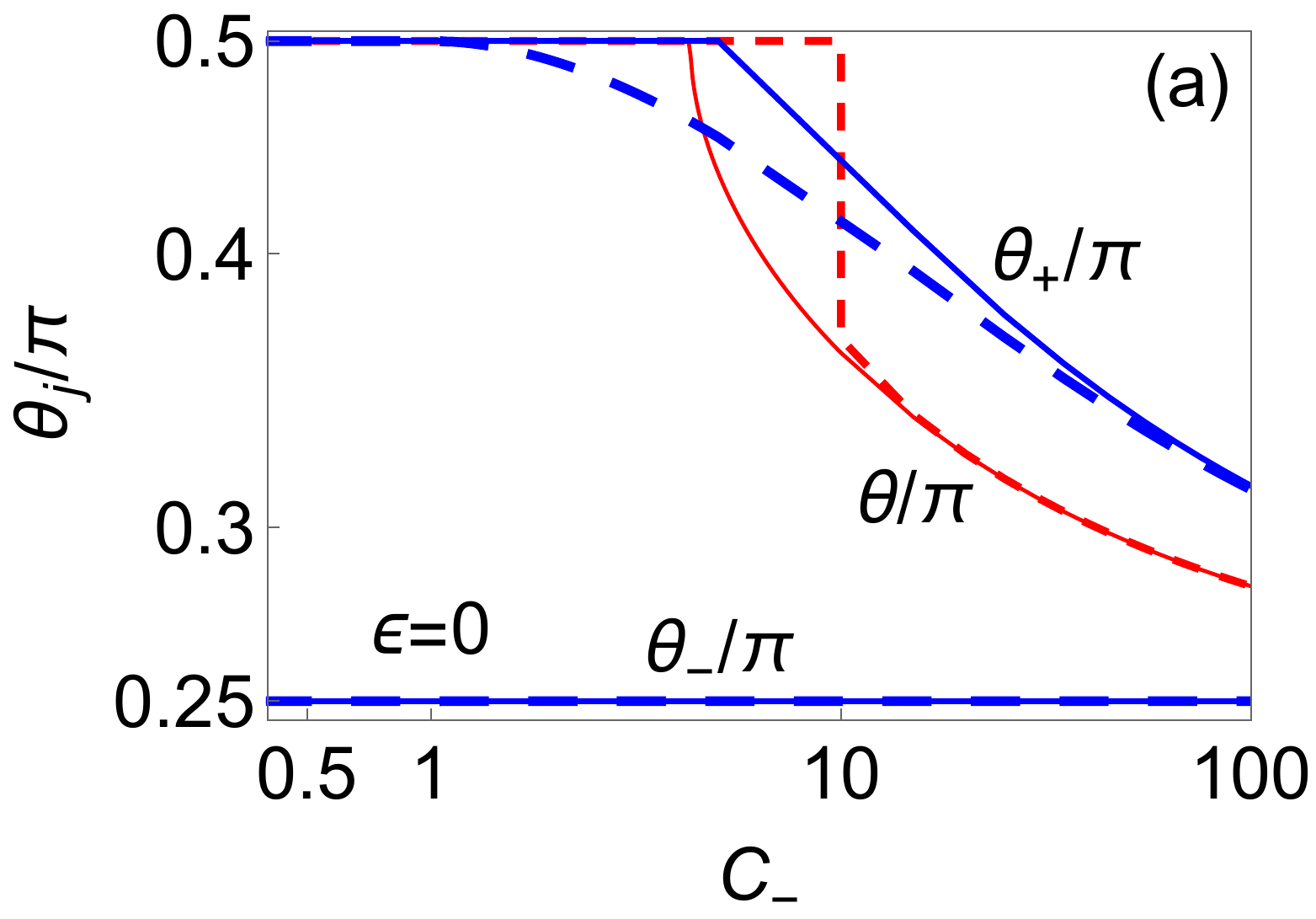}
\includegraphics[width=0.49\columnwidth]{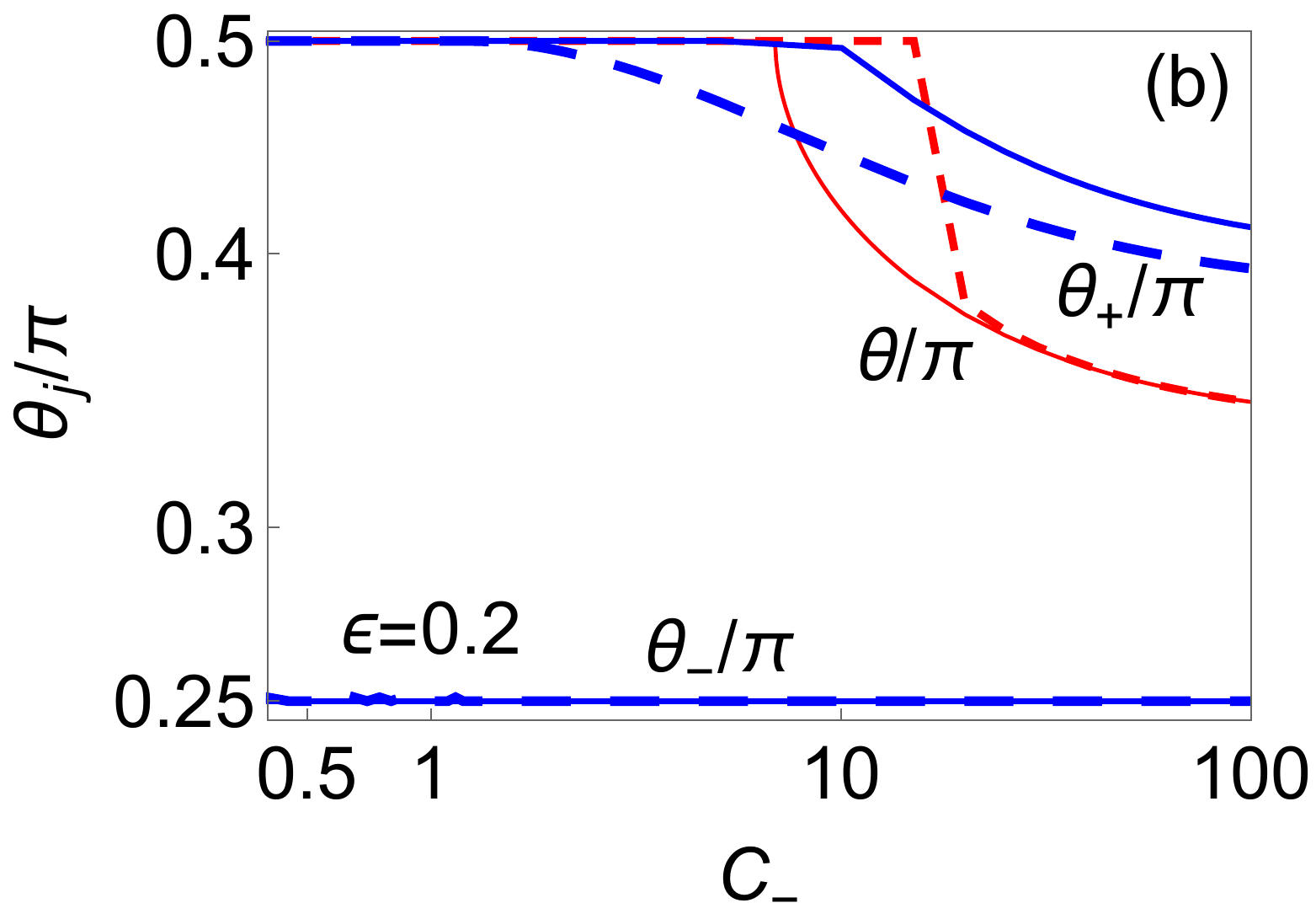}
\includegraphics[width=0.49\columnwidth]{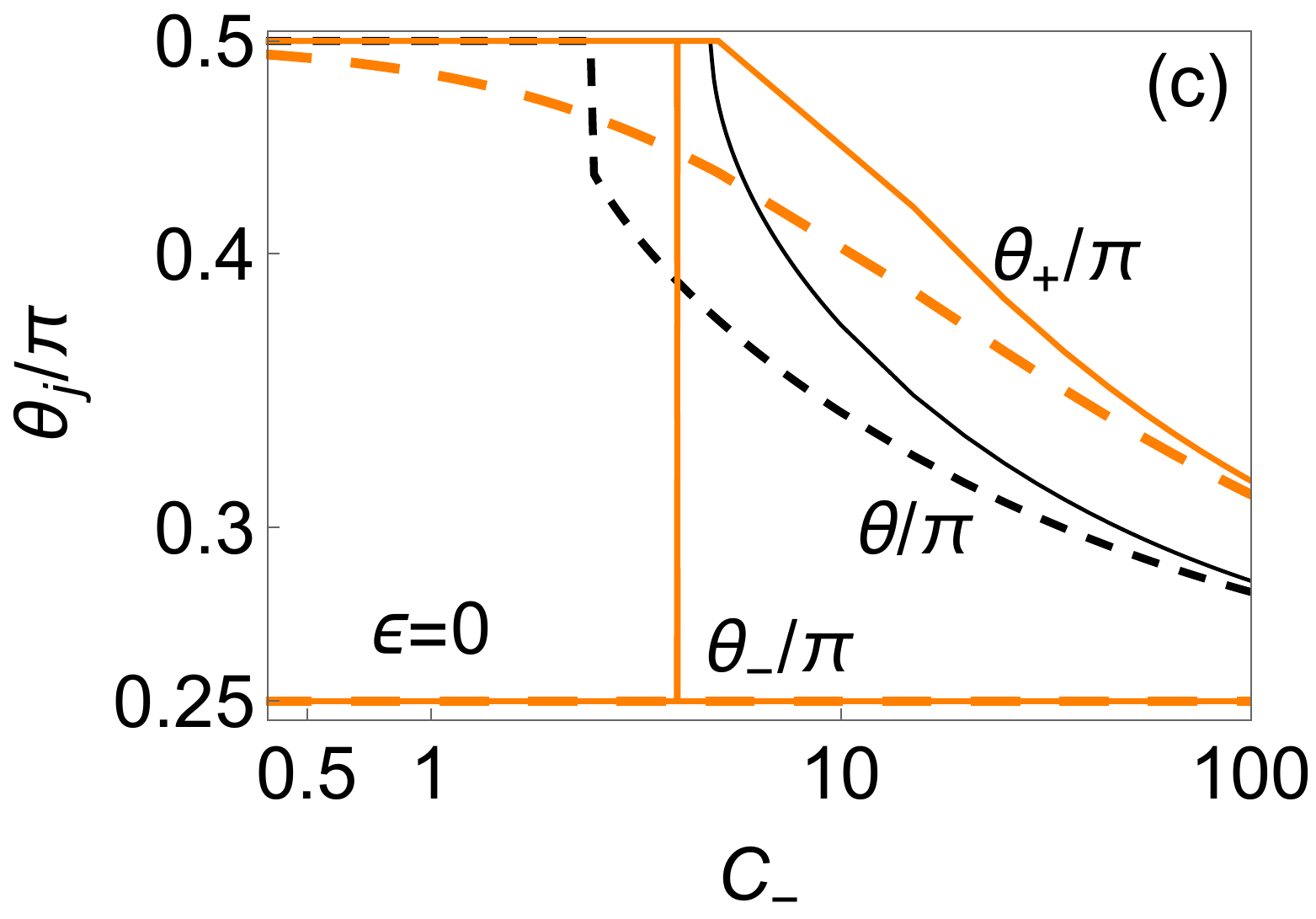}
\includegraphics[width=0.49\columnwidth]{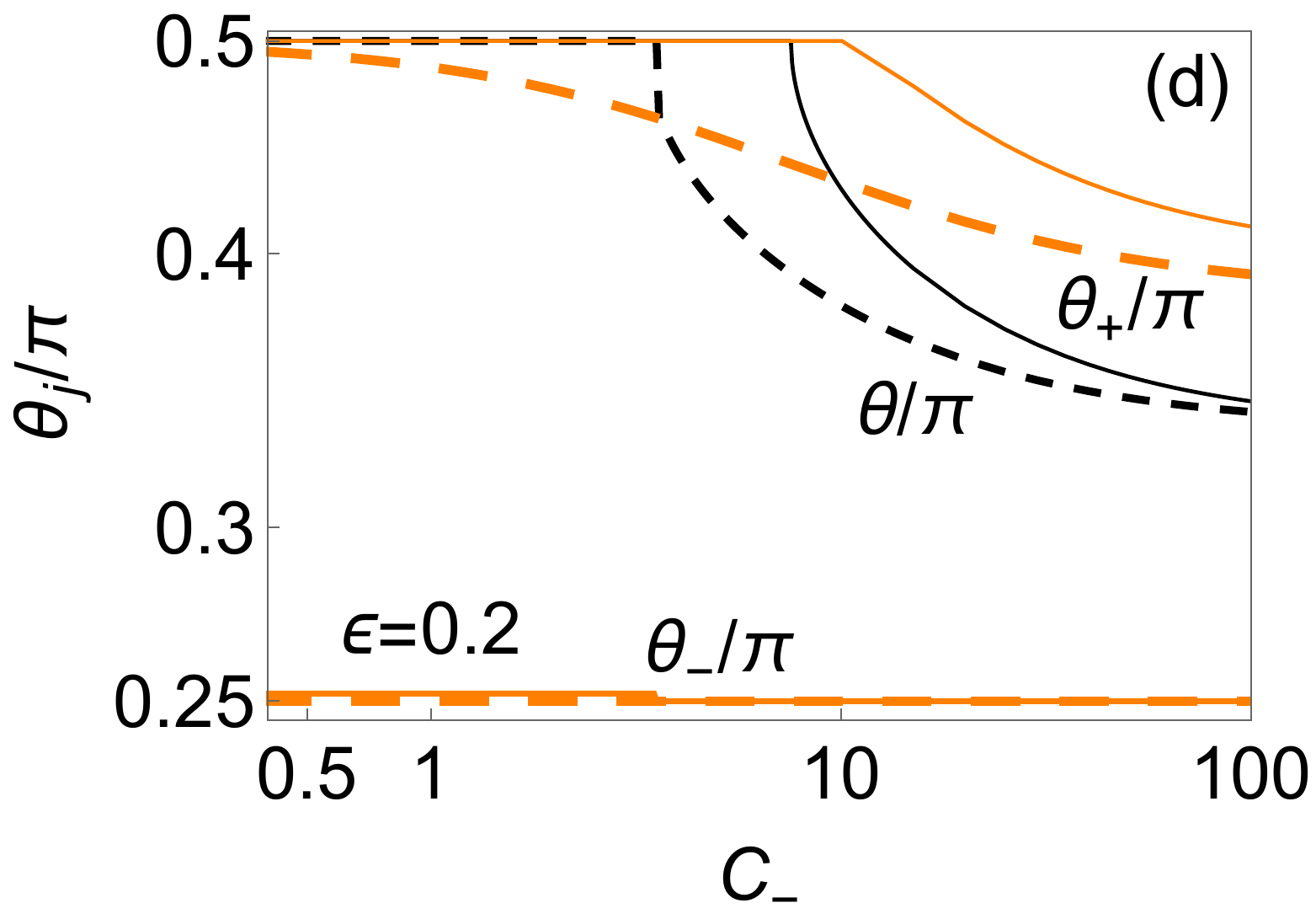}}
\caption{The optimal interaction angles $(\theta_{\pm})$  for the minimized $E_{+ \mid -}$ (dashed) and $E_{- \mid +}$ (solid) as a function of $C_{-}$ for the fixed thermal parameters as intrinsic linewidth $\gamma_{\pm,0}=2\pi \times0.1\text{Hz}$ and thermal occupation number $\bar{n}_{\pm}=10^{5}$ when $f=0$ (thin red curves) and $f \neq 0$ (thick blue curves), as shown in panels (a) and (b). Panels (c) and (d) represent the scenario of asymmetric thermal parameters with intrinsic linewidths $\gamma_{-,0}=2\pi \times0.1\text{Hz}$, $\gamma_{+,0}=2\pi \times20\text{kHz}$ and thermal occupation numbers $\bar{n}_{-}=10^{5}$, $\bar{n}_{+}=0$, when $f=0$ (thin black curves) and $f \neq 0$ (thick orange curves).  Note that in panel (c) when minimizing $E_{-\mid+}$ (solid), the optimal values are $\theta_{-}=\pi/2$ and $\theta_{+}=\pi/4$ when $C_{-}$ is small, and then the optimal values of $\theta_{\pm}$ switch, i.e., $\theta_{-}=\pi/4$ and $\pi/4<\theta_{+}\leq\pi/2$ with increasing $C_{-}$.}
\label{fig: opt_angle}
\end{figure}

As illustrated in Fig.~\ref{fig: opt_angle}, the optimal asymmetric interaction shows $\theta_{-}=\pi/4$ when minimizing $E_{+\mid-}$ and $E_{-\mid+}$ below $1/2$. The optimal $\pi/4<\theta_{+}\leq\pi/2$ indicates that more beam-splitter interaction (cooling) than parameteric-down-conversion interaction (heating) is required for the second $(+)$ subsystem. The optimal asymmetric interaction corresponds to $f<0$, which is consistent with our analysis in the main text. 
In the case of the symmetric coupling $(\theta_{\pm}=\theta)$, i.e., $f=0$, $\pi/4<\theta<\theta_{+}$ indicates that to generate EPR steering, less cooling for the second $(+)$ subsystem is required, but the first $(-)$ subsystem should also implement the same degree of the dynamical cooling.

The effect of the optical transmission losses ($\epsilon$) on optimizing the interaction angles can be seen by comparing Figs.~\ref{fig: opt_angle}(a) and \ref{fig: opt_angle}(b), and also Figs.~\ref{fig: opt_angle}(c) and \ref{fig: opt_angle}(d). It shows that the optimal interaction of the first $(-)$ subsystem is still  $\theta_{-}=\pi/4$, but more beam-splitter interaction is required for the second $(+)$ subsystem in the presence of the transmission loss $\epsilon \neq 0$. Moreover, the optimal choices of $\theta_{+}$ for minimizing $E_{+\mid-}$ and $E_{-\mid+}$ differ even in the limit of large cooperativities $C_{\pm}$, which is different with the case of the symmetric coupling  when $\epsilon \neq 0$.

\begin{figure}[!h]
\centering{\includegraphics[width=0.48\columnwidth]{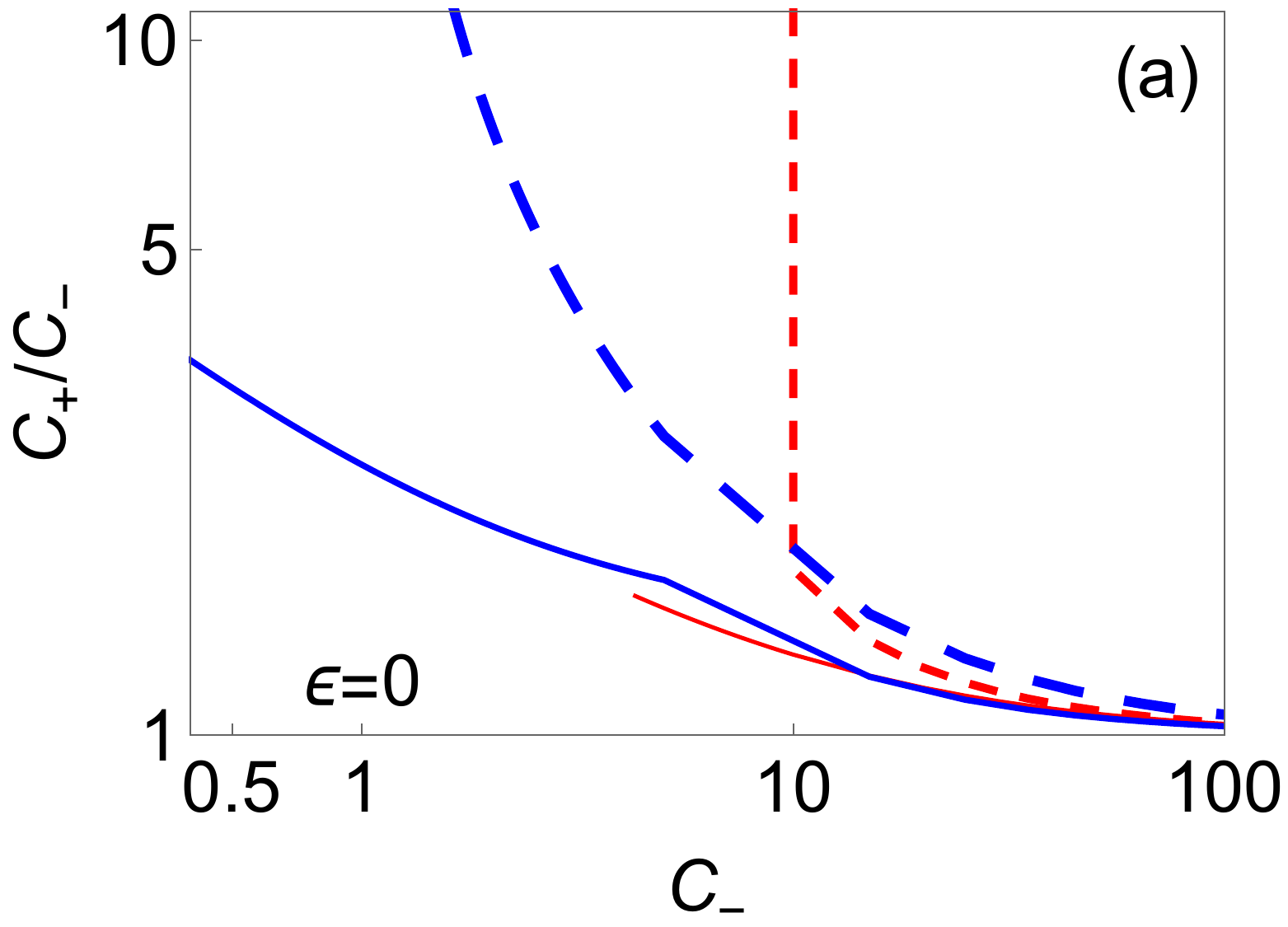}
\includegraphics[width=0.48\columnwidth]{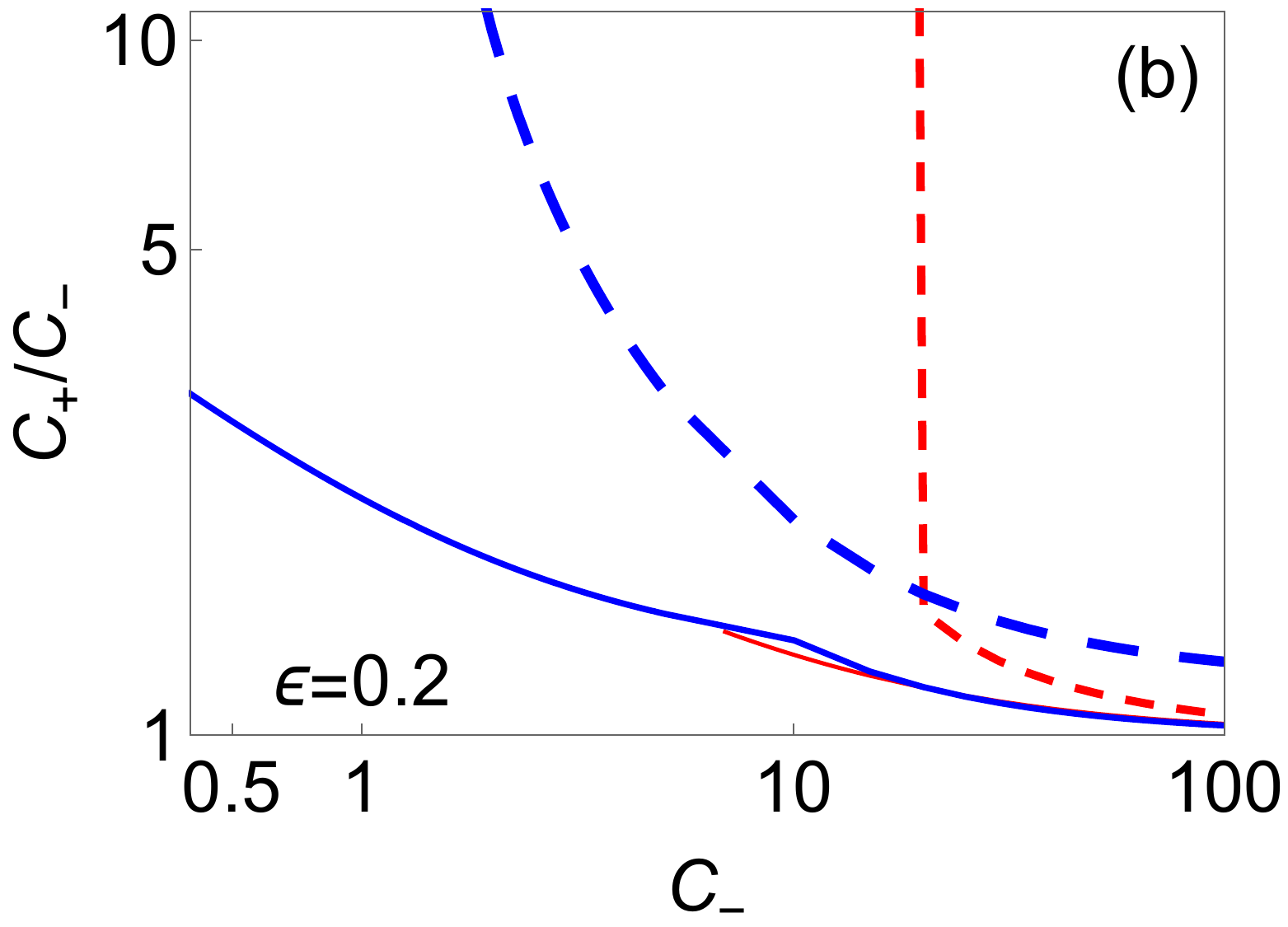}
\includegraphics[width=0.49\columnwidth]{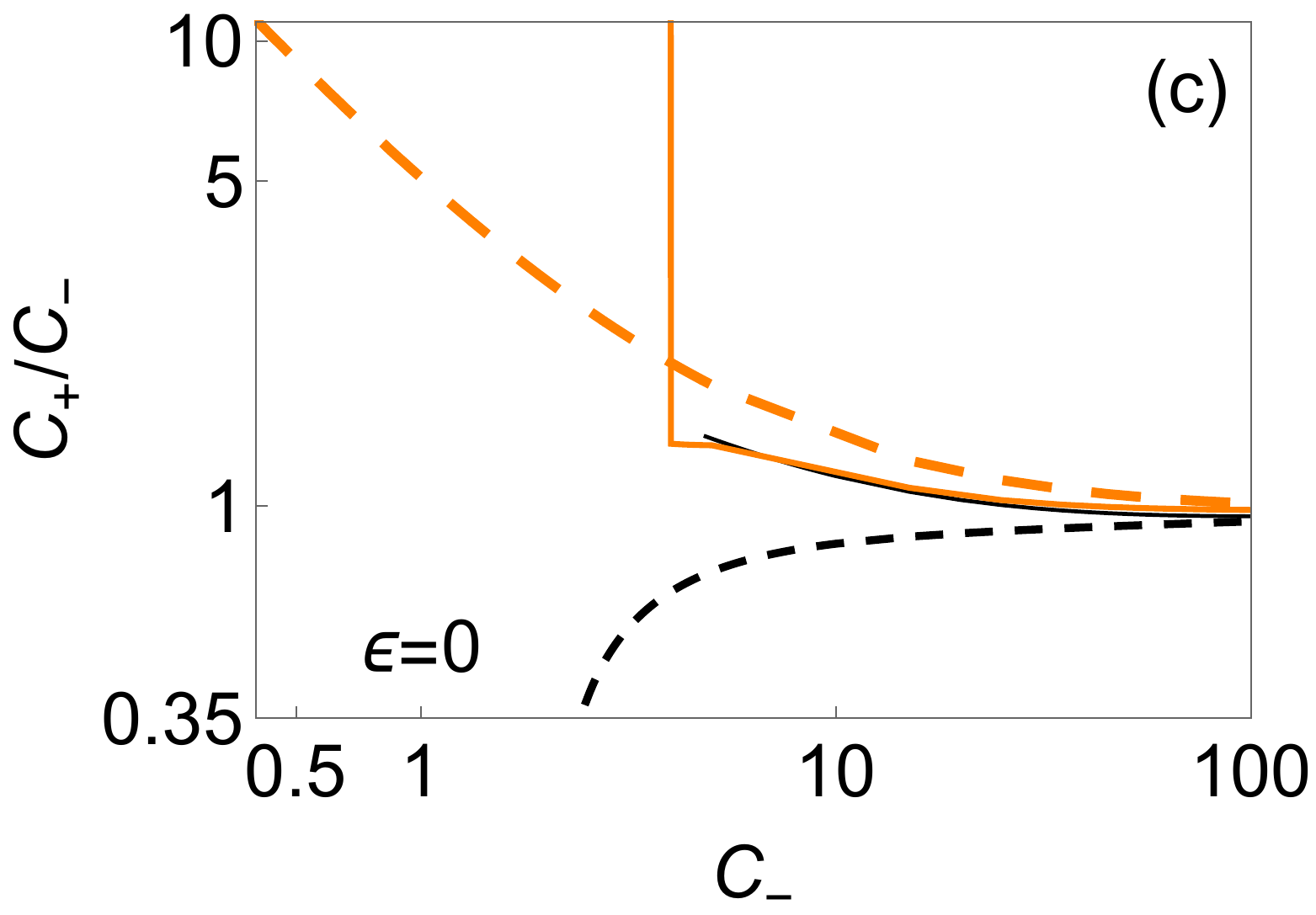}
\includegraphics[width=0.49\columnwidth]{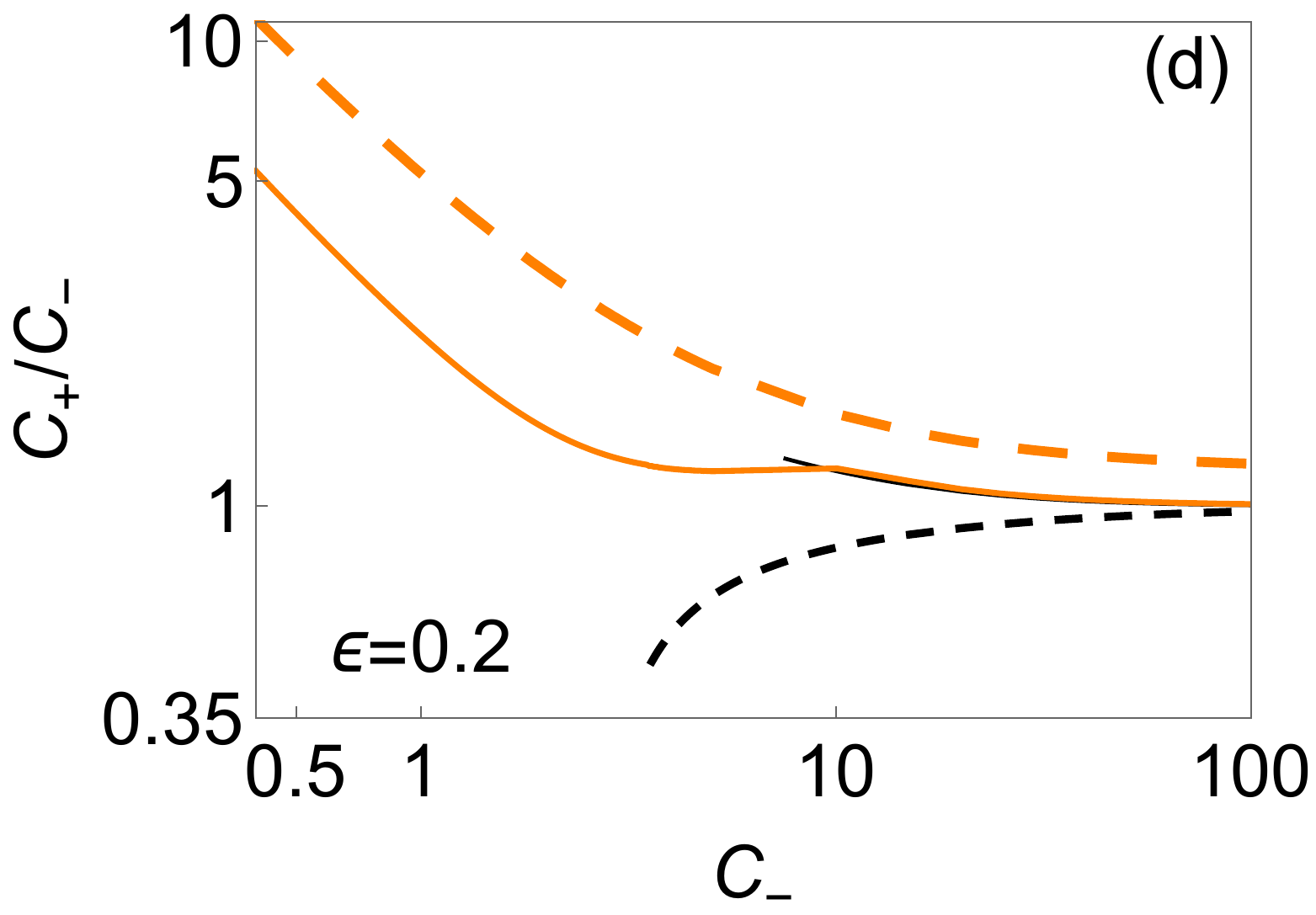}}
\caption{The optimal ratio of quantum cooperativities $(C_{+}/C_{-})$ for the minimized $E_{+ \mid -}$ (dashed) and $E_{- \mid +}$ (solid) as a function of $C_{-}$ for the fixed thermal parameters as intrinsic linewidth $\gamma_{\pm,0}=2\pi \times0.1\text{Hz}$, thermal occupation number $\bar{n}_{\pm}=10^{5}$ when $f=0$ (thin red curves) and $f \neq 0$ (thick blue curves), as shown in panels (a) and (b). Panels (c) and (d) represent the scenario of asymmetric thermal parameters with intrinsic linewidths $\gamma_{-,0}=2\pi \times0.1\text{Hz}$, $\gamma_{+,0}=2\pi \times20\text{kHz}$ and thermal occupation numbers $\bar{n}_{-}=10^{5}$, $\bar{n}_{+}=0$, when $f=0$ (thin black curves) and $f \neq 0$ (thick orange curves).}
\label{fig: opt_c}
\end{figure}

Now we come to the results of the optimal ratio of the quantum cooperativities $C_{+}/C_{-}$ for the minimized $E_{+\mid-}$ and $E_{-\mid+}$. As shown in Fig.~\ref{fig: opt_c}, the optimal $C_{+}/C_{-}$ to minimize the steering parameters when allowing asymmetric coupling are larger than 1, while in the case of the symmetric coupling, the optimal $C_{+}/C_{-}<1$ occurs when minimizing $E_{+\mid-}$ for the fixed asymmetric thermal parameters (black dashed curve). 

In the limit of the large cooperativity $C_{-}$, the optimal $C_{+}/C_{-}$ for the two minimized steerabilities ($E_{\pm\mid\mp}$) both approach 1 both for the cases of symmetric and asymmetric coupling, in the absence of optical transmission loss $\epsilon=0$. However, in the presence of $\epsilon$, the optimal $C_{+}/C_{-}$ that minimize $E_{+\mid-}$ and $E_{-\mid+}$ remain different when choosing the asymmetric coupling. 

Note that in Fig.~\ref{fig: opt_c} the optimal ratio $C_{+}/C_{-}$ for the symmetric coupling to minimize $E_{-\mid+}$ (red solid and black solid curves) is plotted in the regime of $\theta<\pi/2$~(Fig.~\ref{fig: opt_angle}). The reason is when the optimal interaction angle is $\theta=\pi/2$, $E_{-\mid+}=(\Delta\hat{X}_{-})^2$  is independent of the value of $C_{+}$, as can be seen from Eqs.~\eqref{eq:ss_EPRvar} and \eqref{eq:E} in the main text. Moreover, when considering the asymmetric thermal parameters, the optimal ratio $C_{+}/C_{-}$ for the symmetric coupling to minimize $E_{+\mid-}$ [black dashed curve in Fig.~\ref{fig: opt_c}(c) and \ref{fig: opt_c}(d)] is plotted in the range where the corresponding optimal interaction angle $\theta<\pi/2$. This can be understood similarly from Eqs.~\eqref{eq:ss_EPRvar} and \eqref{eq:E} in the main text, where $E_{+\mid-}=(\Delta\hat{X}_{+0})^2=(\Gamma_{+}/2+\tilde{\gamma}_{+,0})/(\gamma_{+,0}+\Gamma_{+})$ when $\theta=\pi/2$. For $\bar{n}_{+}=0$, $E_{+\mid-}=1/2$ becomes irrelevant with the value of $C_{+}$ and also $\gamma_{+,0}$.

\section{Comparison between unconditional and conditional schemes}\label{app:Cond-comparison}

\begin{figure}[!h]
\centering{\includegraphics[width=0.49\columnwidth]{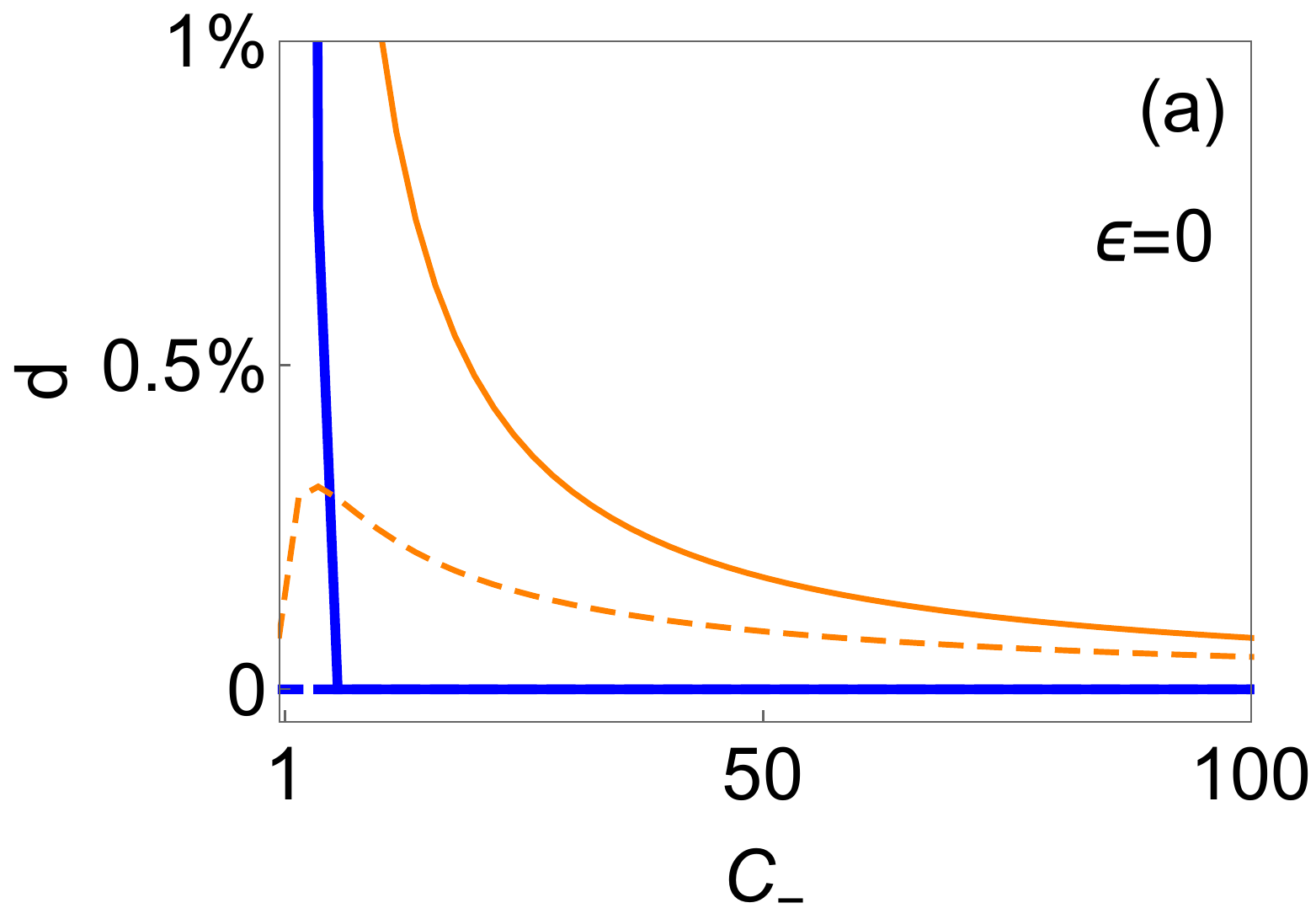}
\includegraphics[width=0.49\columnwidth]{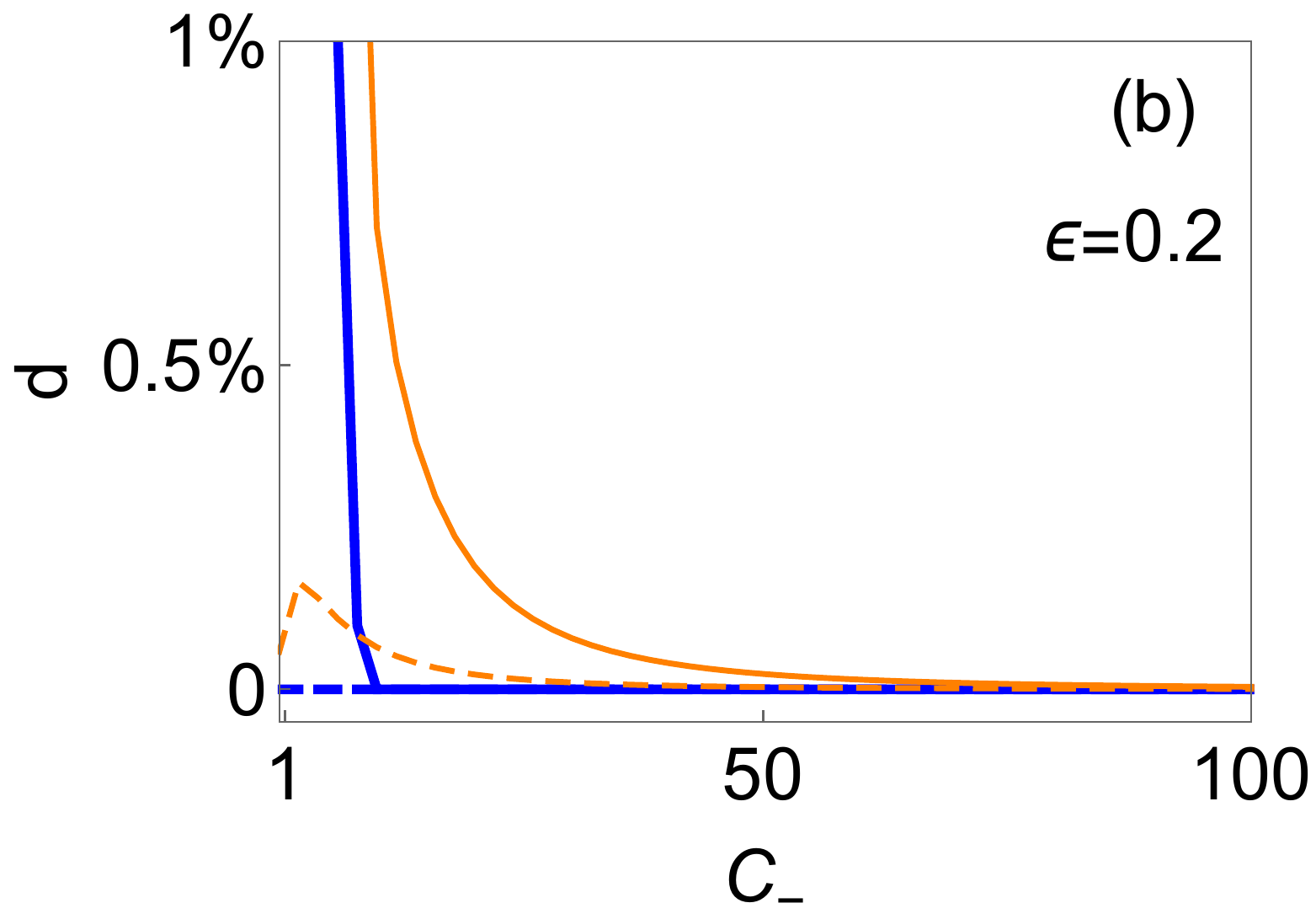}}
\caption{The relative improvement $d$ of the optimal $E_{+ \mid -}$ (dashed) and $E_{- \mid +}$ (solid) as a function of $C_{-}$ for the fixed thermal parameters as intrinsic linewidth $\gamma_{\pm,0}=2\pi \times0.1\text{Hz}$ and thermal occupation number $\bar{n}_{\pm}=10^{5}$ (thick blue curves). The thin orange curves are plotted based on the asymmetric thermal parameters: intrinsic linewidths $\gamma_{-,0}=2\pi \times0.1\text{Hz}$, $\gamma_{+,0}=2\pi \times20\text{kHz}$ and thermal occupation numbers $\bar{n}_{-}=10^{5}$, $\bar{n}_{+}=0$.}
\label{fig: d}
\end{figure}
We use the relative improvement of the minimized steering parameters ($E_{+\mid-}$, $E_{-\mid+}$) of the conditional scheme over the unconditional scheme to compare the performances of the two schemes in the dynamically stable regime, i.e., $\gamma_{\pm}>0$. The relative improvement is defined as $d=(E_{\text{u}}-E_{\text{c}})/E_{\text{u}}$, where $E_{\text{u/c}}$ is the minimal degree of the steering parameters that can be achieved by optimizing the relevant parameters in the unconditional and conditional schemes.

As shown in Fig.~\ref{fig: d}, the relative improvement of the optimized steerabilities ($E_{-\mid+}$, $E_{+\mid-}$) by adding the measurement of the joint output light field is small for the cases of typical interest. In particular, for the client-to-server steerability $E_{+|-}$, $d<1\%$  within the entire range of interest for $C_{-}$. We hence conclude for the generation of EPR steering that our unconditional scheme can achieve practically the same optimal performance as that of the conditional scheme under the conditions considered in Fig.~\ref{fig: opt_E} in the main text.  
\end{appendix}

%%%%%%%%
%\bibliographystyle{unsrt}
\bibliography{Steering}
\end{document}